\documentclass[twocolumn,showpacs,preprintnumbers,amsmath,amssymb,
floatfix,nofootinbib]{revtex4}
\usepackage{amsmath,amsfonts,bbm,euscript,hyperref}
\usepackage{graphicx}% Include figure files
\usepackage{dcolumn}% Align table columns on decimal point
\usepackage{bm}% bold math
\usepackage{epsfig}
\epsfclipon

\newcommand{\nco}{\newcommand}
\nco{\beq}{\begin{equation}} \nco{\eeq}{\end{equation}}
\nco{\beqa}{\begin{eqnarray}} \nco{\eeqa}{\end{eqnarray}}
\def\be{\begin{equation}}
\def\ee{\end{equation}}    
\def\baray{\begin{eqnarray}}
\def\earay{\end{eqnarray}}

\nco{\lra}{\leftrightarrow}

\nco{\sss}{\scriptscriptstyle} \nco{\dphi}{\varphi}
\nco{\lsim}{\mbox{\raisebox{-.6ex}{~$\stackrel{<}{\sim}$~}}}
\nco{\gsim}{\mbox{\raisebox{-.6ex}{~$\stackrel{>}{\sim}$~}}}

\def\IK{\relax{\rm I\kern-.20em K}}
\def\IM{\relax{\rm I\kern-.20em M}}

\def\subo{{(1)}}
\def\subt{{(2)}}
\def\subth{{(3)}}

\def\lsim{\mbox{\raisebox{-.6ex}{~$\stackrel{<}{\sim}$~}}}
\def\gsim{\mbox{\raisebox{-.6ex}{~$\stackrel{>}{\sim}$~}}}
\def\sss{\scriptscriptstyle}

\def\done{\delta^{(1)}}
\def\dtwo{\delta^{(2)}}
\def\sH{\mathcal{H}}
\def\Mpl{M_p}
\def\Lap{\partial^{k}\partial_{k}}
\def\Linv{\triangle^{-1}}
\def\Linvd{\triangle^{-2}}
\def\sR{\mathcal{R}}
\def\Grad{\vec{\nabla}}

\begin{document}

%\preprint{McGill 04-xxx}

\title{Nongaussian and nonscale-invariant perturbations from tachyonic preheating
in hybrid inflation}

\author{Neil Barnaby, James M.\ Cline}

\affiliation{%
\centerline{Physics Department, McGill University,
3600 University Street, Montr\'eal, Qu\'ebec, Canada H3A 2T8}
e-mail: barnaby@hep.physics.mcgill.ca, jcline@physics.mcgill.ca }

\date{January, 2006}

\begin{abstract}   

We show that in hybrid inflation it is possible to generate large
second-order perturbations in the cosmic  microwave background due to
the instability of the tachyonic  field during preheating. We
carefully calculate this effect from the tachyon contribution  to the
gauge-invariant curvature perturbation, clarifying some confusion in
the literature concerning nonlocal terms in the tachyon curvature
perturbation; we show explicitly that such terms are absent.  We
quantitatively compute the nongaussianity generated by the tachyon 
field during the preheating phase and translate the experimental
constraints on the  nonlinearity parameter $f_{NL}$ into constraints
on the parameters of the model. We also show that nonscale-invariant 
second-order perturbations from the tachyon field with spectral index
$n=4$ can become larger than the inflaton-generated first-order perturbations,
leading to stronger constraints than
those coming from nongaussianity.  The width of the
excluded region in terms of the logarithm of 
the dimensionless coupling $g$, grows linearly with 
the log of the ratio of the Planck mass to the tachyon VEV, 
$\log(M_p/v)$; hence very large regions are ruled out if the
inflationary scale $v$ is small.  We apply these results to
string-theoretic brane-antibrane inflation, and find 
a stringent upper bound on the string coupling, $g_s < 10^{-4.5}$.

\end{abstract}

% PACS, the Physics and Astronomy
                             % Classification Scheme.
%\keywords{Suggested keywords}%Use showkeys class option if keyword
                              %display desired
\maketitle

\section{Introduction}

Inflation \cite{Guth} has become the dominant paradigm for the
generation of the anisotropies of the Cosmic  Microwave Background
(CMB) (see \cite{RiottoReview,BrandenbergerReview} for a review). 
During inflation quantum fluctuations are redshifted out of the
horizon where they become ``frozen'' and later re-enter the horizon
after Big Bang Nucleosynthesis.  Observation of the temperature
anisotropies of the CMB directly constrain the curvature perturbation
somewhat before cosmological scales re-enter the horizon.  Current
observations show that the curvature perturbation is gaussian with a
scale-invariant spectrum to within  experimental limits, which is
consistent with the predictions of inflation.  Nevertheless, some 
nongaussianity is expected to be generated during inflation
\cite{inflationaryNG}-\cite{otherNG}  (see \cite{NGreview} for a
review).  It is interesting to study the level of nongaussianity,
usually parameterized by a dimensionless nonlinearity parameter
$f_{NL}$, in various scenarios for inflation, since it can be an
important way to discriminate between different models.  For example,
the curvaton mechanism \cite{curvaton} for the generation of
cosmological perturbations may be  ruled out by future
non-observation of nongaussianity \cite{curvatonNG} (see, however, 
\cite{curvatonNGexception}).

One way to generate significant levels of nongaussianity at the end
of  inflation is by preheating \cite{preheatNG0}-\cite{preheatNG6}. 
Preheating refers to the  nonperturbative production of particles  
which arises either due to coherent oscillations \cite{preheating} or
tachyonic (spinodal) instability \cite{tachyonic1}-\cite{tachyonic3}
(see also \cite{ReheatTheory,Bassett}).
It has also been suggested that preheating might affect the first
order variation of the  curvature perturbation, perhaps dominating
the contribution of the inflaton and thus modifying the standard
predictions of inflation \cite{firstorderreheating}-\cite{FB2}.  

In this paper we carefully reconsider the generation of nongaussian
fluctuations from tachyonic preheating in hybrid inflation
\cite{hybrid} using second order cosmological perturbation theory
similarly to \cite{Acquaviva,EV}.  See \cite{doubleinflation} for
constraints on the parameter space of hybrid inflation coming from the
WMAP data.  During hybrid inflation the
inflaton field $\varphi$ is displaced from its preferred value, and
inflation is driven by the false vacuum energy of the ``waterfall''
or tachyon field $\sigma$ which is trapped in its false  vacuum by
interactions with the inflaton.  When the inflaton reaches some
critical value the tachyon  effective mass squared  becomes negative
and its fluctuations grow exponentially due to the spinodal
instability.  The exponential growth continues until the fluctuations
of the tachyon start to oscillate about the true vaccuum.  At this
stage the tachyon fluctuations become nonperturbative and their
back-reaction brings inflation to an end.  Here we study the
evolution of the second-order curvature perturbation due to the
tachyonic instability and find that it can lead to a large level of
nongaussianity, if the tachyon was  lighter than the Hubble scale
$m_\sigma < 3H/2$, during a sufficient part of the inflationary epoch.
 (If $m_\sigma \gg H$, the fluctuations of $\sigma$ are exponentially 
suppressed during
inflation.)  In the hybrid inflation model, it can be natural to
have $m_\sigma < 3H/2$ during inflation, since $m_\sigma^2$ changes
sign at the end of inflation.  If the inflaton is rolling slowly
enough, $m_\sigma^2$ will naturally already be close to zero even 
before the end of inflation, at the time of horizon crossing.

The scenario is summarized as follows.  During inflation the squared
mass of the tachyon field varies linearly with the number of
e-foldings of inflation, in the vicinity of the point where it passes
through zero: $m^2_\sigma = c H^2 N$, taking $N=0$ to be the value
where $m^2_\sigma=0$, and $c$ is a function of the parameters of
hybrid inflation.    Inflation starts at some $N\equiv N_i < 0$, and
ends at $N\equiv N_* > 0$, for a total of  $N_e = -N_i + N_*$
e-foldings.  We are going to limit our inquiry to regions of
parameter space where the linear behavior of  $m^2_\sigma$ is valid
during the full duration of inflation, since this technical
assumption makes the calculations tractable; however we will show
that this assumption must often be satisfied anyway, as a consequence of
the experimental limit on the inflationary spectral index.  At $N=0$,
fluctuations of the tachyon field start to grow exponentially, and
can make a large contribution to the curvature perturbation at second
order in cosmological perturbation theory, resulting in a
scale-noninvariant component both in the CMB  temperature power
spectrum, and its nongaussian correlations.   Our goal is to derive
constraints on the parameter space of hybrid inflation from these
effects.

We start in section \ref{II} by defining the perturbations up to
second order in the metric and inflationary fields.  In section
\ref{III} the hybrid inflation model is reviewed, starting with the
dynamics of the fields at zeroth order, and then their first order
perturbations, with emphasis on the dynamics of the tachyon field
fluctuations toward the end of inflation.
In section \ref{IV} we solve for the second
order curvature perturbation which is induced by the tachyonic
fluctuations. This is used in section \ref{V} to compute the
bispectrum (three-point function) and the nonlinearity
parameter $f_{NL}$, as well as contributions to the spectrum
itself. In section \ref{VI} we incorporate experimental constraints
on  nongaussianity as well as on the
spectrum, to derive excluded regions in the parameter space of
the hybrid inflation model. 
In section \ref{VII} we consider the possibility of generating
significant nongaussianity in brane inflation.
We conclude in section \ref{VIII},
comparing our results with previous treatments.  We give technical
details concerning mode functions in de Sitter space, the second order
perturbed Einstein equations, the inflaton curvature perturbation, 
Fourier transforms of convolutions, the construction of the tachyon
curvature perturbation and the approximation of an adiabatically
varying tachyon mass in appendices A-F.

\section{Metric and Matter Perturbations}
\label{II}

In this section we write down the perturbations of the metric and
matter fields about a spatially flat  Robertson-Walker background
following \cite{Acquaviva}.  Greek indices run over the full
spacetime  $\mu,\nu = 0,1,2,3$ while latin indices run only over the
spatial directions $i,j=1,2,3$.  We will work mostly  in conformal
time $\tau$, related to cosmic time $t$ by $dt = a d\tau$.  
Differentiation with
respect to conformal time will be denoted by
$f' = \partial_{\tau}f$ and with respect to cosmic time by $\dot{f} =
\partial_t f$.

The metric is expanded up to second order in fluctuations as
\begin{eqnarray}
  g_{00} &=& -a(\tau)^2 \left[ 1 + 2 \phi^\subo + \phi^\subt   \right] \\
  g_{0i} &=& a(\tau)^2 \left[ \partial_i \omega^\subo + \frac{1}{2} \partial_i \omega^\subt 
                               + \omega_i^\subt  \right] \\
  g_{ij} &=& a(\tau)^2 \left[ (1 - 2 \psi^\subo - \psi^\subt) \delta_{ij} 
                              + D_{ij} (\chi^\subo + \frac{1}{2}\chi^\subt) \right.\nonumber \\
         &+& \left. \frac{1}{2}(\partial_i \chi_j^\subt +\partial_j \chi_i^\subt  + \chi_{ij}^\subt) \right]
\end{eqnarray}
where $D_{ij} = \partial_i \partial_j -
\frac{1}{3}\delta_{ij}\partial^k \partial_k$ is a trace-free 
operator.  The fluctuations are decomposed such that the vector
perturbations are transverse  $\partial^i \omega_i^\subt = \partial^i
\chi_i^\subt = 0$ while the tensor perturbations are transverse, 
traceless and symmetric: $\partial^i \chi_{ij}^\subt = 0$,
$\chi^{i\subt}_{i} = 0$,  $\chi_{ij}^\subt = \chi_{ji}^\subt$.  In
the above we have neglected the vector and tensor perturbations at 
first order, which are small, since vector perturbations decay with
time, while tensors are suppressed by the slow roll parameter
$\epsilon$.  The same is not true at second order, however, since
the  second order tensors and vectors are sourced by the first order
scalar perturbations.  We adopt the generalized longitudinal gauge
defined by $\omega^\subo = \omega^\subt = \omega_i^\subt = 0$ and
$\chi^\subo = \chi^\subt = 0$.  The metric in this gauge becomes
\begin{eqnarray}
  g_{00} &=& -a(\tau)^2 \left[ 1 + 2 \phi^\subo + \phi^\subt   \right] \\
  g_{0i} &=& 0 \\
  g_{ij} &=& a(\tau)^2 \left[ (1 - 2 \psi^\subo - \psi^\subt)\, \delta_{ij} 
            \right.\nonumber\\ &+& \left.\frac{1}{2}(\partial_i \chi_j^\subt +\partial_j \chi_i^\subt  + \chi_{ij}^\subt)   \right].
\end{eqnarray}

In hybrid inflation the matter content consists of
two scalar fields which are expanded in perturbation theory as
\begin{eqnarray}
  \varphi(\tau,\vec{x}) &=& \varphi_0(\tau) + \done \varphi(\tau,\vec{x}) 
                            + \frac{1}{2}\dtwo \varphi(\tau,\vec{x}) \\
  \sigma(\tau,\vec{x}) &=& \sigma_0(\tau) + \done \sigma(\tau,\vec{x})
                           + \frac{1}{2} \dtwo \sigma(\tau,\vec{x}).
\end{eqnarray}
where $\varphi$ is the inflaton and $\sigma$ the tachyon (or
``waterfall'' field).  In hybrid inflation the time-dependent vacuum
expectation value (VEV) of the tachyon field is set to zero
$\sigma_0(\tau)=0$, about which we will say more later.  

The perturbations are defined so that $\langle \delta^{(i)} \varphi
\rangle = 0$, hence $\langle \varphi(\tau,\vec{x}) \rangle =
\varphi_0(\tau)$.  At first  order in perturbation theory this is
automatic since $\done\varphi$ contains only one
annihilation/creation  operator. However, at higher order in
perturbation theory the homogeneous $k=0$ mode of the fluctuation
must be subtracted by hand in order to ensure that all of the zero
mode of the field is described by the nonperturbation background. 

The Einstein tensor and stress-energy tensor expanded up to second
order in perturbation theory can be found in \cite{EV}.  We do not
reproduce these results here, but we have carefully checked all the
results from \cite{EV} which are relevant for our analysis.

\section{Hybrid Inflation}
\label{III}

We consider hybrid inflation in which both the inflaton and the
tachyon are real fields with the potential
\begin{equation}
\label{pot}
  V(\varphi,\sigma) = 
\frac{\lambda}{4} \left( \sigma^2 - v^2 \right)^2 
+ \frac{m^2_\varphi}{2}\varphi^2
                      + \frac{g^2}{2} \varphi^2 \sigma^2.
\end{equation}
This potential will give rise to topological defects
 at the end of inflation---domain walls
in the $\sigma$ field---which could produce
large nongaussianities apart from the ones which we consider
(\cite{Robert}).  But these domain walls must somehow be rendered
unstable to avoid overclosure of the universe, for example through
the addition of a small term like  $\mu \sigma^3$ to
(\ref{pot}).   We will ignore this issue here. It would be
interesting to consider a complex tachyon field
\cite{inprogress} whose defects formed at the
end of inflation are cosmic strings, which are much more
phenomenologically viable than domain walls.

\subsection{Background Dynamics}

At the homogeneous level, the usual Friedmann and Klein-Gordon equations 
for the scale factor and the matter fields are
\begin{eqnarray}
  3 H^2 &=& \frac{\kappa^2}{2} \left( \dot{\varphi}_0^2 + \dot{\sigma}_0^2 \right) + \kappa^2 V, 
  \label{friedmann}\\
  0 &=& \ddot{\varphi}_0 + 3 H \dot{\varphi}_0 + \frac{\partial
V}{\partial \varphi} \label{KGzeroth}\\
  0 &=& \ddot{\sigma}_0 + 3 H \dot{\sigma}_0 + \frac{\partial V}{\partial \sigma} \label{KG2zeroth}
\end{eqnarray}
where $\kappa^2 = \Mpl^{-2} = 8\pi G_{N}$.  Here and elsewhere
the  potential and its derivatives are understood to
be evaluated on the background values of the fields so that  $V =
V(\varphi_0,\sigma_0)$ and $\partial V / \partial \varphi =  \partial
V / \partial \varphi |_{\{\varphi = \varphi_0, \sigma = \sigma_0\}}$,
for example.  For the potential (\ref{pot}) we have 
$\partial V / \partial \sigma = \partial^2 V / \partial \sigma
\partial \varphi = 0$, provided that $\sigma_0=0$.
We will apply this  simplification to all subsequent results.

To see why  one should set $\sigma_0(\tau) = 0$, notice that
the tachyon effective mass is $m_{\sigma}^2 = g^2 \varphi_0^2 -
\lambda v^2 > 0$.  That is, the tachyon mass squared starts out
being positive during inflation.  Provided that there was a long
enough prior period of inflation, any initial departure of $\sigma$
from zero would be exponentially damped.  At some point, 
$m_{\sigma}^2$ becomes negative, and the tachyonic instability
begins.  However, it is still true that $\sigma_0(\tau)$ remains
zero even then, since the universe will consist of equal numbers
of domains with $\sigma>0$ and $\sigma<0$.  On average, these give
zero, which is the definition of the zeroth order field $\sigma_0$.
The departures of $\sigma$ from zero between domain walls which form
are taken account in the fluctuations of the field.  Thus it is
consistent to set $\sigma_0(\tau)=0$ for all times in our analysis.
(Notice that the situation would be different if we considered a complex
scalar field $\sigma$ in which case the phase transition would lead to 
cosmic string formation and the radial degree of freedom $|\sigma|$
would not average to zero at late times.)

We will make extensive use of the 
slow roll parameters, defined by
\begin{eqnarray*}
  \epsilon &=& \frac{\kappa^2}{2} \frac{\dot{\varphi}_0^2}{H^2} = -\frac{\dot{H}}{H^2} 
  \cong\frac{\Mpl^2}{2 V^2}\left( \frac{\partial V}{\partial \varphi}\right)^2, \\
  \epsilon - \eta &=& \frac{\ddot{\varphi}_0}{H \dot{\varphi}_0} 
  \cong \epsilon - \frac{\Mpl^2}{V} \left( \frac{\partial^2 V}{\partial \varphi^2}\right)
\label{sr}
\end{eqnarray*}
so that, during inflation
\beq
  \eta \cong 4\, \frac{\Mpl^2\, m^2_\varphi}{\lambda\, v^4}, \qquad
  \epsilon \cong  8\left(\frac{M_p\, m^2_\varphi\, \varphi_0}{\lambda\,
v^4}\right)^2 
\eeq
Notice that if $m^2_\varphi\varphi_0^2\ll \lambda v^4$, then $\epsilon \ll \eta$.
This is equivalent to demanding that the false vacuum energy of the 
tachyon dominates during inflation, which is the assumption usually
made for hybrid inflation: 
\[
  V(\varphi_0, \sigma_0=0) = \frac{\lambda v^4}{4} + \frac{m^2_{\varphi}}{2} \varphi_0^2
                                     \cong \frac{\lambda v^4}{4}.
\]

During the slow roll phase the inflaton equation of motion
$  3 H \dot{\varphi}_0 + m^2_{\varphi} \varphi_0 \cong 0$
has solution
\begin{equation}
\label{inflatonbkg}
  \varphi_0(t) = 
\varphi_s \exp \left( -\frac{m^2_\varphi (t-t_s)}{3H} \right)
	= \varphi_s \left(a(t)\over a_s\right)^{-\eta}
\end{equation}
where we used $a(t) = a_s e^{H(t-t_s)}$, with $t_s$ an arbitrary time.  
The Hubble scale remains approximately constant, $3 H^2 \cong \lambda
v^4 / (4\Mpl^2)$. Since $\varphi_0$ is decreasing and
$\sigma_0=0$, the slow roll 
parameter $\epsilon$ actually decreases slowly
during inflation while $\eta$ remains constant.

\subsection{Inflationary Dynamics of First Order Fluctuations}
\label{firstordersection}

Having argued that the background inflaton field $\varphi_0$ is described
by (\ref{inflatonbkg}) with $H$ approximately constant during 
both inflation and 
the tachyonic instability phase, we now briefly discuss the dynamics
of the first order metric and inflaton  fluctuations for
$\epsilon,\eta \ll 1$.  This section is largely review since when
$\sigma_0=0$ the first order metric and inflaton perturbations obey
exactly the same equations as in single field inflation; however, we
include some details of the calculations since similar equations will
arise when we study the second order metric fluctuations.\footnote{As
will be shown, the second order metric fluctuation $\phi^\subt$ obeys
an inhomogeneous equation where the differential operator is
identical to the one which determines the dynamics of $\phi^\subo$;
thus an understanding of the first order solutions simplifies the
construction of the Green function for the  second order
fluctuations.}\ \ We work in conformal time which is
most convenient for the ensuing calculations. 

The $\done G^{i}_{j} = \kappa^2 \done T^{i}_{j}$ Einstein equation
for $i \not= j$ implies that  $\phi^\subo = \psi^\subo$ which is a
well known result.  We apply this simplification in the following.  
The $\done G^{0}_{i} = \kappa^2 \done T^{0}_{i}$
Einstein equation is a constraint
\begin{equation}
\label{(0,i)}
  \phi'^\subo + \sH \phi^\subo = \frac{\kappa^2}{2} \varphi_0'\, \done \varphi
\end{equation}
which means that the first order metric perturbation $\phi^\subo$ and
the first order inflaton perturbation $\done \varphi$ are not
independent.  Once either $\phi^\subo$ or $\done \varphi$ is known,
the other may be computed from (\ref{(0,i)}), though it is
simplest to solve for $\phi^\subo$ and use
(\ref{(0,i)}) to compute $\done \varphi$.  One obtains a dynamical
equation for the metric perturbation by applying (\ref{(0,i)}) to the
sum of $\done G^{0}_{0} = \kappa^2 \done T^{0}_{0}$ and  $\done
G^{i}_{i} = \kappa^2 \done T^{i}_{i}$ Einstein equations.  The result
is
\begin{equation}
\label{conformalPhi1}
  \phi''^\subo_k + 2 \left( \sH - \frac{\varphi_0''}{\varphi'_0} \right) \phi'^\subo_k 
  + \left[ 2 \left( \sH' - \sH \frac{\varphi''_0}{\varphi'_0} \right) + k^2 \right] \phi^\subo_k = 0.
\end{equation}
Notice that the perturbed Klein-Gordon equation for the inflaton is
not needed to close the system of equations. We discuss the slow roll
solutions of (\ref{conformalPhi1}) in some detail since later on we
will need to construct the Green function for the operator in the
left-hand-side of (\ref{conformalPhi1}).

The conformal time slow roll parameters are
\begin{eqnarray}
  \epsilon &=& 1 - \frac{\mathcal{H}'}{\mathcal{H}^2} 
               = \frac{\kappa^2}{2} \frac{\varphi_0'^2}{\mathcal{H}^2} \label{ce}, \\
  \epsilon - \eta &=& \frac{\varphi_0''}{\mathcal{H} \varphi_0'} - 1 \label{cd}.
\end{eqnarray}
During a pure deSitter phase $\epsilon = \eta = 0$ and the scale
factor evolves as $a(\tau) = -1/(H\tau)$ with $H$ constant.  During
inflation, however, the Hubble scale evolves slowly as (\ref{ce}) so
that for small $\epsilon$ one has \cite{RiottoReview}
\[
  a(\tau) = -\frac{1}{H \tau} \frac{1}{1-\epsilon}
\]
and
\begin{equation}
\label{SRH}
  \sH = \frac{a'}{a} = \dot{a}  = a H = \frac{-1}{\tau (1-\epsilon)}.
\end{equation}
Note that during a slow roll phase we can treat $\epsilon$ and $\eta$
as constant (even though $H$ is not exactly constant) since
$\epsilon'$, $\eta'$ are second order in slow roll parameters:
\[
  \epsilon' = -2 \epsilon (\eta - 2\epsilon) \sH \ll \epsilon \sH, \hspace{5mm}
  \eta' = 2 \epsilon \, \eta \sH \ll \eta \sH.
\]
The above equality for $\eta'$ is only strictly correct if
$\partial^3 V / \partial \varphi^3 = 0$ which is  true in the case of
interest.  The statement that the slow roll parameters are
approximately constant is,  however, quite general.\footnote{Of
course this statement generalizes to cosmic time as well.}

The dynamical equation for $\phi^\subo$ (\ref{conformalPhi1}) can be
rewritten in terms of the slow  roll parameters as
\begin{equation}
\label{SRphi1}
  \phi''^\subo_k - \frac{2}{\tau} (\eta -\epsilon) \phi'^\subo_k 
  + \left[ \frac{2}{\tau^2} (\eta - 2 \epsilon) + k^2 \right] \phi^\subo_k = 0
\end{equation}
where we used (\ref{SRH}) and dropped higher order terms in
$\epsilon$, $\eta$.  Treating the slow roll parameters as constant,
the equation (\ref{SRphi1}) has an exact solution
\begin{equation}
\label{SRphi1soln}
  \phi^\subo_k(\tau) = (-\tau)^{1/2 + \eta - \epsilon} 
    \left[c_1(k) H_{\nu}^\subo(-k\tau) + c_2(k) H_{\nu}^\subt(-k\tau) \right]
\end{equation}
where $\nu \cong 1/2 + (3\epsilon - \eta) $ to lowest order in slow
roll parameters.

It remains to fix the coefficients $c_1(k)$, $c_2(k)$ in
(\ref{SRphi1soln}).  The variable in terms of which the action is
canonically normalized is the Mukhanov variable
\cite{BrandenbergerReview}
\[
  V_k^\subo = a \left[ \done \varphi_k + \frac{\varphi'_0}{\sH} \phi^\subo_k \right]
\]
where the inflaton fluctuation is solved for using the constraint
equation (\ref{(0,i)}). We fix $c_1(k)$, $c_2(k)$ by requiring that
$V_k^\subo \cong e^{-i k \tau} / \sqrt{2 k}$ on small scales  $-k
\tau \gg 1$ which corresponds to the usual Bunch-Davies vacuum
choice.  This leads to
\[
  c_1(k) = i \frac{H}{\Mpl} \sqrt{\frac{\pi \epsilon}{2}} \,
           \frac{\exp\left[\frac{i \pi}{2}(\nu + 1/2) \right]}{2 k}, 
           \hspace{5mm} c_2(k) = 0.
\]
The solution for the metric perturbation, then, is
\beqa
\label{infPhi1}
  \phi^\subo_k(\tau) &=& i \sqrt{\frac{\pi \epsilon}{2}} \frac{H}{\Mpl} 
  \frac{\exp \left[ \frac{i \pi}{2} ( \nu + 1/2 )\right] }{2 k}
\nonumber\\
&\times&  (-\tau)^{1/2 + \eta -\epsilon} H^\subo_{\nu} (- k \tau).
\eeqa
It is straightforward to construct the inflaton fluctuation using
(\ref{(0,i)}).  One may also compute the comoving curvature
perturbation at first order, $\sR^\subo_k = \sH V_k^\subo / (a
\varphi'_0)$, and verify  that the solution (\ref{infPhi1}) reproduces
the usual inflationary prediction for the power spectrum.  One may also
verify that in the limit $\epsilon \rightarrow 0$ and $\eta \rightarrow 0$
the analysis of this subsection reproduces deSitter mode functions, which
are discussed in appendix A.

\subsection{Conditions for a slowly varying tachyon mass}

Now we come to an important point for this paper, that if $\eta$
is sufficiently small, then the tachyon is also a light field 
during some part of the observable period of inflation.  
The tachyon mass is given
by $m^2_\sigma = -\lambda v^2 + g^2 \varphi_0^2(t)$. If we choose the
arbitrary time $t_s$ in (\ref{inflatonbkg}) to be when
$m^2_\sigma =0$, then $g^2 \varphi_s^2 = \lambda v^2$, and
\beqa
\label{mseq}
	m^2_\sigma &=& -\lambda v^2\left(1 - 
	\left(a(t)\over a_s\right)^{-2\eta}\right)
\cong -2\eta\lambda v^2 H (t-t_s) \nonumber\\
	&=& -2\eta\lambda v^2\, N
\eeqa
where $N$ is the number of e-foldings of inflation occurring after
the tachyonic instability begins.  At some maximum value $N=N_*$,
inflation will end.  If the inflaton rolls slowly
enough, then the tachyon mass remains close to zero for a significant
number of e-foldings.  In general,
we will have $-N_i\equiv N_e-N_*$ e-foldings of inflation before the spinodal
time, followed by $N_*$ e-foldings during the preheating phase.

The approximation in (\ref{mseq}), 
that the tachyonic mass changes slowly enough for
its time dependence to be approximated as linear, is true so long as
$|2\eta N| \ll 1$.  This can be rephrased using the definition of
$\eta$ in (\ref{sr}), and eliminating $m^2_\varphi$ using the COBE
normalization of the inflationary power spectrum: ${V/(M_p^4\epsilon)} \cong 150\pi^2(2\times 10^{-5})^2
= 6\times 10^{-7}$ (see for example ref.\ \cite{LL}).
Using $V = \frac14\lambda v^4$ and eq.\ (\ref{sr}) for $\epsilon$,
the COBE normalization gives
\beq
\label{cobe}
	{m^2_\varphi} \cong 230\, g\lambda\, 
	{v^5\over M_p^3}
\eeq
Then  with $N\sim 60$, the requirement
$|2\eta N| < 1$ becomes
\beq
	g < 10^{-5} \, {M_p\over v}
\label{lincon}
\eeq

Interestingly, the bound (\ref{lincon}) turns out to be a requirement
that must often be satisfied for different reasons, namely the
experimental limit on the spectral index of the first-order inflaton
fluctuations. In terms of the slow-roll parameters, the deviation of the
spectral index from unity is given by
\[
  n-1 = 2\eta - 6\epsilon \cong 2 \eta
\]
where we have used the fact that $\epsilon \ll \eta$ in hybrid
inflation  (this is equivalent to the requirement that the energy
density which drives inflation is dominated by $\lambda v^4/4$).  The
experimental constraint on the spectral index is roughly $|n-1| \lsim
10^{-1}$.  Writing $\eta$ in terms of model parameters this
translates into the constraint 
\beq
g\, \frac{v}{\Mpl} \lsim 5\times 10^{-5}
\label{etacon}
\eeq
This is just five times weaker than the technical assumption (\ref{lincon}).

In passing, we note that 
there is also a lower bound on $g$ from the assumption that the
false vacuum energy density is dominated by $\lambda v^4/4 >
m_\phi^2\varphi_0^2/2$.  Using   $g^2 \varphi_s^2 = \lambda v^2$ and
(\ref{cobe}), one finds
\beq
\label{glb}
	g > 460\, \lambda\, {v^3\over M_p^3}
\eeq

\subsection{Tachyonic Instability}
\label{sec:tp}

To quantify the evolution of the tachyonic instability at the end of
inflation, we consider the equation of motion for the tachyon field
fluctuation in Fourier space,
\begin{equation}
\label{tachyonEOM}
  \done \ddot{\sigma}_k + 3 H \done \dot{\sigma}_k + 
  \left[ \frac{k^2}{a^2} + \left( g^2 \varphi_0^2 - \lambda v^2\right) 
\right] \done \sigma_k = 0.
\end{equation}
Once $\varphi_0 < \lambda^{1/2} v / g$ the tachyon effective mass
parameter  $\partial^2 V / \partial \sigma^2$ becomes negative and
the fluctuations $\done \sigma$ are amplified due to the spinodal
instability.  The efficient transfer of energy from the false vacuum
energy  $\lambda v^4 /4$ to the fluctuations $\done \sigma$ is
refered to as \emph{tachyonic preheating} in the literature
\cite{tachyonic1}-\cite{tachyonic3}.

The initial studies of tachyonic preheating focused on the flat space
dynamics of $\done \sigma_k$ in the  instantaneous quench
approximation.  In this approximation the field $\done \sigma$ is
initially assumed to  have zero mass and at $t=0$ 
 a negative mass squared term  $-|m_\sigma^2| (\done \sigma)^2 /2$ is turned on.  We briefly review these
dynamics here following closely \cite{tachyonic1}.  Initially the
tachyon has the usual Minkowski space mode functions  $e^{-ikt + i
\vec{k}\cdot\vec{x}} / \sqrt{2k}$ \footnote{Even in deSitter space
the Bunch-Davies vacuum choice will ensure that this behaviour is
respected on small scales $k \gg a H$ regardless of the tachyon mass
during inflation. See appendix A for  a review.}. Once the negative
mass squared term is turned on the modes  with $k = |\vec{k}| <
|m_\sigma|$
grow exponentially with a dispersion 
\beqa
\label{ht}
 \left\langle(\done
\sigma)^2\right\rangle &=& \frac{1}{4\pi^2} \int_0^{|m_\sigma|} dk\, k\,
e^{2t\sqrt{|m_\sigma^2| - k^2}}\\
&=& {|m_\sigma^2|\over 4\pi^2\alpha^2}
\Big(e^\alpha(\alpha-1)+1\Big);\quad\alpha \equiv 2|m_\sigma|t
\nonumber
\eeqa
which produces a spectrum with an
effective cutoff $k_{\mathrm{max}} = |m_\sigma|$.  The tachyonic growth
persists until the dispersion saturates at the value
\beq
\label{end}
\left\langle(\done \sigma)^2\right\rangle^{1/2} \cong {v\over 2}
\eeq
at which point the curvature of the
effective potential vanishes and the tachyonic growth is replaced by
oscillations about the true vacuum.  This process completes within a
time 
\begin{equation} \label{spinodaltime} t_s \sim
\frac{1}{2|m_{\sigma}|}\ln\left(\frac{\pi^2}{\lambda}\right) 
\end{equation}
which we call the spinodal time.  At this point a large fraction of
the vacuum energy $\lambda v^4 / 4$ has been converted into gradient
energy of the field $\done \sigma$ so that the universe is divided
into domains with $\left\langle(\done \sigma)^2\right\rangle^{1/2} =
\pm v$ of size $l \sim |m_{\sigma}|^{-1}$ and on average one still has
$\left\langle\sigma\right\rangle = 0$ so that $\sigma_0(t) = 0$. 
These analytical argument  are backed up by semi-classical lattice
field theory simulations in \cite{tachyonic1,tachyonic2}.

The discussion of the dynamics of tachyonic preheating above apply
strictly only in Minkowski space.  The dynamics of tachyonic 
preheating including the dynamics of the inflaton but neglecting
the expansion of the universe were considered in \cite{tachyonic3}.
The dynamics of tachyonic preheating including both the dynamics
of the inflaton and the expansion of the universe were considered
in \cite{falsevacuumdecay} wherein the authors reach conclusions 
identical to those discussed above.  The authors of 
\cite{falsevacuumdecay} also find that the spinodal time is somewhat 
modified from (\ref{spinodaltime}) due to the background
dynamics (see also \cite{falsevacuumdecay2,falsevacuumdecay3}).  

Notice that we do
not need to replace the average of the fluctuations 
$\left\langle(\done \sigma)^2\right\rangle^{1/2}$ with an effective
homogeneous background $\sigma_0(\tau)$ because we work to second
order in perturbation theory and the effect of these fluctuations
enters into the calculation through the second order perturbed energy
momentum tensor $\left\langle\dtwo T^{\mu}_{\nu}\right\rangle$.
When $\left\langle\dtwo T^{\mu}_{\nu}\right\rangle$ becomes sufficiently
large the backreaction will stop inflation.  We take (\ref{end}) as our criterion
for the end of inflation.  We have checked numerically that this is a somewhat
more stringent constraint than demanding that the energy density in the fluctuations
$\done \sigma$ does not dominate over the false vacuum energy which drives
inflation $\lambda v^4 / 4$.

In the present work, we are interested in a situation which is different
from the instantaneous quench, where the instability may turn on slowly compared to the Hubble
expansion, rather than suddenly.  We are approximating the time
dependence of the tachyon mass as being linear around the time when
it vanishes, eq.\ (\ref{mseq}), 
 so the mode equation can be written
in the form 
\begin{equation}
\label{tEOM2}
 {d^2\over dN^2} \done{\sigma}_k + 3 {d\over dN}  \done{\sigma}_k + 
  \left[{\hat k^2}e^{-2N} - c N \right] \done \sigma_k = 0.
\end{equation}
where $N=H(t-t_s)$, 
$\hat k \equiv k/H$ and, incorporating the COBE normalization
as in (\ref{cobe}),
\beq
\label{ceq}
	 c \cong 22000\, g\, M_p / v\,.
\eeq  
From eq.\ (\ref{lincon}), $c$ is limited to values
\beq
	c\ll \left(M_p\over v\right)^2
\eeq
for the validity of the approximation that the tachyon mass squared varies
linearly with time.
The quantum mechanical solution in terms of annihilation and creation
operators $a_k,\ a_k^\dagger$ has the usual form
\footnote{Our conventions for fourier transforms and mode functions are discussed
in detail in appendix D.}
\beq
\label{quantum}
  \done \sigma(x) = \int {d^{\,3}k\over (2\pi)^{3/2}}
	\,a_k \, \xi_k(N)\, e^{ikx}+ {\rm h.c.}
\eeq
but the mode functions $\xi_k$ will be complicated by the
time-dependence of the tachyon mass.  We normalize the mode functions
$\xi_k$ according to the usual Bunch-Davies prescription which is discussed
in subsection \ref{firstordersection} and also in appendix A.

\begin{figure}[htbp]
\bigskip \centerline{\epsfxsize=0.5\textwidth\epsfbox{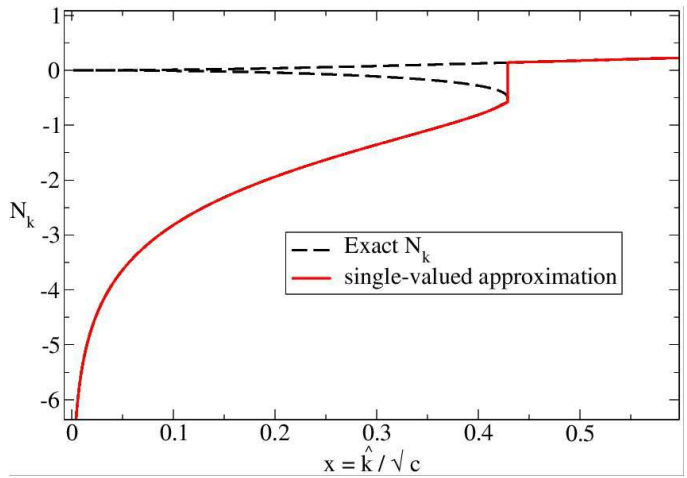}}
%\begin{verse}
%\vskip-0.25cm
\caption{Exact solution and our approximation for the function
$N_k$ in eq.\ (\ref{Nk}).
}
\label{figN}
\end{figure}

Since (\ref{tEOM2}) has no closed-form solution, we approximate it in
two regions.  First, when ${\hat k^2}e^{-2N} >  c |N|$, we ignore the
mass term and use the massless solutions, $\xi_k \sim 
a^{-3/2} H_{3/2}^{(1)}(\hat k e^{-N})$.  We match this onto the
solution in the region where ${\hat k^2}e^{-2N} <  c |N|$, where the
term ${\hat k^2}e^{-2N}$ is ignored in the equation of motion.
The transition between the two regions occurs at different times 
$N_k$ for different wavelengths, given implicitly by
\beq
\label{Nk}
	N_k = 	\ln {\hat k\over\sqrt{c}} - \ln\sqrt{|N_k|} 
\eeq
This is a multivalued function of $x\equiv \hat k/\sqrt{c}$, because
for $x < (2 e)^{-1/2}$, the $\hat k^2$ term in the differential
equation comes to dominate again for a short period around 
$N=0$, the moment when the tachyon is massless.  To deal with this,
we are going to assume that the solution is still well-approximated by
the massive one during this short period.  This amounts to 
replacing the multivalued function with the single-valued one
shown in figure \ref{figN}.   We checked this approximation
in conjunction with other approximations we will make
for the mode functions, as described below eq.\ (\ref{zk}).

In the second region, with $N>N_k$, the solutions are approximated by
Airy functions, but it is more convenient to use the WKB approximation
to obtain an expression in terms of elementary functions. 
Ignoring overall phases, in this way we obtain 
\beq
\label{mode}
	\xi_k \cong \left\{  \begin{array}{ll}
	\left({2 H\hat k^3}\right)^{-1/2} %e^{i\hat k\, e^{-N}}
	\left(1 + i \hat k e^{-N}\right), & N < N_k \\
	b_k\,{ e^{-\frac32 N + \frac{9}{4c}z^{3/2}}(1+|z|)^{-1/4} }, &  N > N_k
\end{array} \right.
\eeq
with
\beq
\label{bk}
	b_k = {1-i\sqrt{c|N_k|}\over \sqrt{2 H}(c|N_k|)^{3/4}}
	{(1 + |z_k|)^{1/4} \over 
	\exp\left({9\over 4c}z_k^{3/2}\right)}
\eeq
and
\beq
	z \equiv \left(1+\frac49 cN\right);
	\quad z_k \equiv \left(1+\frac49 cN_k\right)
\label{zk}
\eeq
In the above expressions, we have for simplicity 
matched the amplitudes but not derivatives of the solutions at
$N=N_k$.   This will not affect the estimates we make below.
We used (\ref{Nk}) to reexpress exponential dependence
on $N_k$ as power law dependence.  Notice that exponent in (\ref{bk})
becomes purely imaginary when $\frac49 {c}N_k < -1$.  We also
replaced $|z|^{1/4} \to (1+|z|)^{1/4}$ to correct the spurious
singularity at $z=0$ where the WKB approximation breaks down.  We
numerically verified that this gives a good approximation to the
exact Airy function solutions.  

\begin{figure}[htbp]
\bigskip \centerline{\epsfxsize=0.5\textwidth\epsfbox{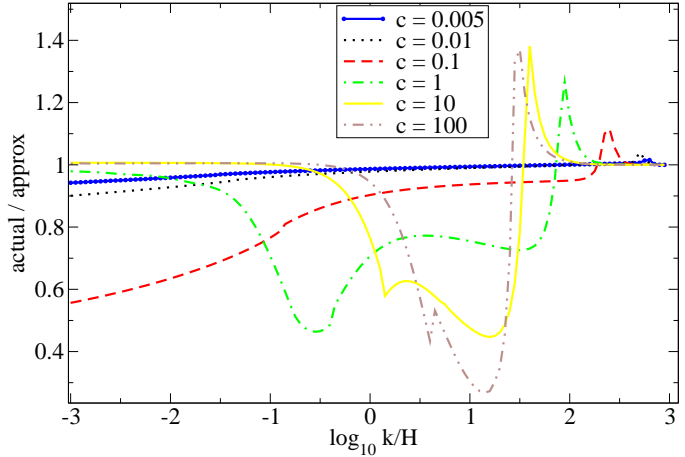}}
%\begin{verse}
%\vskip-0.25cm
\caption{Ratio of the exact 
mode functions to the approximation
(\ref{mode}), at large times.}
\label{correction}
\end{figure}

Moreover, we have checked the
approximate solution by numerically integrating the mode equations,
starting from the small-$N$ region $k^2 e^{-2N}\gg c|N|$, 
where the massless solutions with known amplitude tell us the
initial conditions, and integrating into the large-$N$ region where
the exponential growth due to the tachyonic instability becomes
important.  We did this for two orthogonal solutions to the mode
equations, $\xi_{1,2}$, whose behavior in the small-$N$ region
is
\beqa
	\xi_1 &=& (2Hk^3)^{-1}e^{-N}\left(k\cos(ke^{-N})-\sin(ke^{-N})
\right)\nonumber\\
\quad \xi_2 &=& (2Hk^3)^{-1}e^{-N}\left(k\sin(ke^{-N})+\cos(ke^{-N})
\right)\nonumber
\eeqa
Evolving these initial conditions to large $N$, the envelope of these
functions, which is also the modulus of the complex solutions, is
$\xi=\sqrt{\xi_1^2+\xi_2^2}$.  We compared this numerical solution to
the modulus $|\xi_k|$ of  our approximation (\ref{mode}) for a large
range of $c$ and $k$ values.  In the large $N$-region, $|\xi_k|$
agrees with $\xi$ up to a numerical factor of order unity.   This
factor, the ratio of the actual solution to
the  approximation, is
shown in figure (\ref{correction}). Because the exponential growth
of the mode function is a very steep function of $N$, these small
errors have an imperceptible effect on the exclusion plots we
will present in section \ref{VI}.  Furthermore, we have checked
which values of $c$ and $k$ actually give constraints in the
parameter space of the hybrid inflation model below, and found that
in the regions where $c$ is large, $k$ is exponentially small.  
Extrapolating the results of figure (\ref{correction}) indicates
that the error becomes quite small as $k\to 0$.  
Therefore our approximations
for the mode functions are quite good for the purposes of this
paper. 

\begin{figure}[htbp]
\bigskip \centerline{\epsfxsize=0.5\textwidth\epsfbox{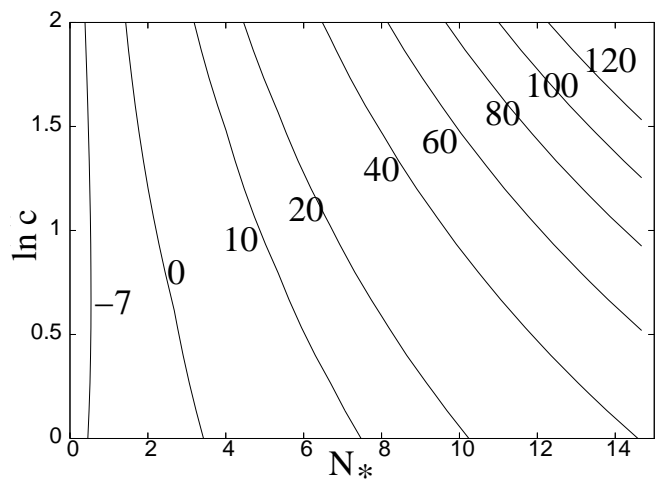}}
%\begin{verse}
%\vskip-0.25cm
\caption{Contours of $\ln M_p^2/\lambda v^2$ in the plane of $N_*$ and
$\ln c$.}
\label{figNs}
\end{figure}

With the above approximate solution, we are in a position to
recompute the dispersion of the tachyon fluctuations in the case
of a more slowly varying tachyon mass, $\langle (\done\sigma)^2\rangle
= (2\pi)^{-3}\int d^3 k |\xi_k|^2$.  Following the discussion of subsection
\ref{sec:tp}, we set this equal to $v^2/4$ at the end of inflation, $N=N_*$:
\beq
\left.\int {d^3 k\over(2\pi)^{3}}  |\xi_k|^2\right|_{N=N_*} = 
{v^2\over 4}
\label{vend}
\eeq
which implicitly determines
$N_*$ in terms of parameters of the hybrid inflation model,
\beq
N_* = N_*\left(g,\lambda,v/M_p\right)
\eeq
The main contribution to the
integral at $N=N_*$
comes from wave numbers for which the exponentially growing
solution in (\ref{mode}) applies.  These modes satisfy $k < k_{\rm
max}\equiv H e^{N_*} \sqrt{c N_*}$.  We have numerically performed
the integral for a wide range of values of $c$ and $N_*$. The
result is displayed in figure \ref{figNs}, where
contours of $\ln M_p^2/\lambda v^2$ are shown in the 
plane of $N_*$ and $\ln c$.  Recall that $c = 22000 
\,g\, (M_p/ v)$, eq.\ (\ref{ceq}).

\section{Second Order Fluctuations in the Long Wavelength Approximation}
\label{IV}
\subsection{The Master Equation}

The authors of \cite{EV} have derived a ``master equation'' for the second order potential 
$\phi^\subt$ which can be written as
\beqa
\label{master}
    \phi''^\subt + 2 \sH (\eta-\epsilon) \phi'^\subt &+&
  \left[ 2 \sH^2(\eta-2\epsilon) - \Lap \right] \phi^\subt\nonumber\\ 
 &=& J(\tau,\vec{x})
\eeqa
where the source terms are constructed entirely from first order quantities and can be split into
inflaton and tachyon contributions
\[
  J(\tau,\vec{x})= J^{\sigma}(\tau,\vec{x}) + J^{\varphi}(\tau,\vec{x}).
\]
Although we have explicitly inserted the slow roll parameters,
equation (\ref{master}) is quite general and we have not yet assumed
that $\eta,\epsilon$ are small (although recall that we have 
set $\sigma_0 = 0$).  We have verified both the second 
order Einstein equations and the master equation presented in
\cite{EV} and these results are discussed in appendix B.

Because the equation (\ref{master}) is linear we can split the
solutions $\phi^\subt$ into three parts: the  solution to the
homogeneous equation, the particular solution due to the inflaton 
souce and the particular solution due to the tachyon source.  
The solution to the homogeneous equation will be proportional to
$\phi^\subo$ since the differential operator on the left hand
side of (\ref{master}) is identical to the operator which determines
$\phi^\subo$, equation (\ref{conformalPhi1}).  In cosmological perturbation
theory the split between background quantities and fluctuations is 
unambigiuous, since background quantities depend only on time while
fluctuations depend on both position and time.  However, the split
between first order and second order fluctuations is arbitrary and
the freedom to include in the solution for $\phi^\subt$ a 
contribution which is proportional to $\phi^\subo$ reflects this.  
We fix this ambiguity by including only the particular solutions for 
$\phi^\subt$ which is due to the source, $J(\tau,\vec{x})$.

During the preheating phase the tachyon  fluctuations are  amplified by a
factor $v/H$ and so $J^{\sigma}$  will come to dominate.  This implies that
the particular solution for $\phi^\subt$ which is due to $J^\sigma$ will come to
dominate over the particular solution which is due to $J^\varphi$.  Although
our analysis will focus on this part of the solution, we will also
consider the inflaton source in order to verify that our formalism
reproduces previous results.

%Notice that the differential operator on the left hand side of
%(\ref{master}) is identical to the operator for the dynamical
%equation for the first order metric fluctuation $\phi^\subo$,
%eq.\  (\ref{conformalPhi1}).  This means that the homogeneous
%solution of (\ref{master}) will be proportional  to (\ref{infPhi1})
%and we may use the analysis of subsection
%\ref{firstordersection} to construct the  Green function for this
%operator.

\subsection{The Green Function}

In this subsection we construct the Green function for the master
equation so that $\phi^\subt$ may be determined in terms of first
order quantities.  As discussed previously we consider only the 
particular solution for
$\phi^\subt$ due to  the source $J$ and neglect the solution to the
homogeneous equation, which is equivalent to assuming that the 
second order fluctuations are zero before the source is turned on. 
During a quasi-deSitter phase the  master equation can be written as
\beqa
\label{samaster}
  \partial_{\tau}\left[ (-\tau)^{2(\epsilon-\eta)}\partial_{\tau}
\phi^\subt_k \right] &+& 
  (-\tau)^{2(\epsilon - \eta)} \left[ \frac{2}{\tau^2} (\eta - 2\epsilon) + k^2\right] \phi^\subt_k
\nonumber\\  &=& (-\tau)^{2(\epsilon-\eta)} J_k(\tau)
\eeqa
where it is assumed that $\epsilon,\eta \ll 1$, and 
the differential operator on the left hand side
of the master equation is written in a manifestly self-adjoint form.  
In deriving (\ref{samaster}) we used (\ref{SRH}) and treated the 
slow roll parameter as constant, which
is consistent at first order in the 
slow roll expansion.  The causal Green 
function for this operator is
\begin{eqnarray}
  G_k(\tau,\tau') &=& \frac{\pi}{2} 
                    \Theta(\tau-\tau') \left( \tau \tau' \right)^{1/2 + \eta - \epsilon}  \label{green} \\
                  &\times& \left[ J_{\nu}(-k\tau) Y_{\nu}(-k\tau') - J_{\nu}(-k\tau') Y_{\nu}(-k\tau)\right]
                  \nonumber
\end{eqnarray}
where the order of the Bessel functions is $\nu \cong 1/2 + 3\epsilon - \eta$ as in subsection 
\ref{firstordersection}.  
%This result is valid only
%when $\epsilon,\eta \ll 1$.  

The solution for the metric perturbation $\phi^\subt$ can then be written as
\begin{equation}
\label{particular}
  \phi^\subt_k(\tau) = \int_{-(1+\epsilon)/a_i H}^{0} d\tau' 
                       G_{k}(\tau,\tau') \left(- \tau' \right)^{2(\epsilon - \eta)} J_{k}(\tau')
\end{equation}
where $a_i = a(t_i)$ is the scale factor at the some initial time, well before
the tachyonic instability has set in.  This solution
is quite general; it applies during any slow roll phase, including
during the tachyonic instability. 

There are several interesting limiting cases of (\ref{green}).
In the long wavelength limit $k \rightarrow 0$ the Green function reduces to
\beqa
  G_{k}(\tau,\tau') &=& \Theta(\tau-\tau')  (1 + 2\eta - 6\epsilon)
\nonumber\\                      
&& \!\!\!\!\!\!\!\!\!\!\!\!\!
\left[  -(-\tau)^{1 + 2\epsilon} (-\tau')^{2(\eta-2\epsilon)} + 
                      (-\tau')^{1 + 2\epsilon} (-\tau)^{2(\eta-2\epsilon)}  \right].
\nonumber
\eeqa
In the case of pure deSitter expansion $\epsilon = \eta = 0$, but for all $k$, the Green function reduces
to
\[
  G_{k}(\tau,\tau') = \Theta(\tau-\tau') \frac{1}{k} \sin\left[k\left(\tau-\tau'\right)\right].
\]
Finally, the form of the Green function in the case of $\epsilon = \eta = 0$ and $k \rightarrow 0$
may be of some interest
\begin{equation}
\label{simple_green}
  G_{k}(\tau,\tau') = \Theta(\tau-\tau')(\tau-\tau').
\end{equation}

\subsection{The Gauge Invariant Curvature Perturbation}

The curvature perturbation is expanded to second order as
\[
  \zeta = \zeta^\subo + \frac{1}{2}\zeta^\subt
\]
In hybrid inflation, where $\sigma_0 = 0$, the first order contribution comes entirely from the inflaton
sector
\[
  \zeta^\subo = -\phi^\subo - \sH \frac{\done \rho}{\rho'_0}
\]
where $\rho_0 = -(T^{\,0}_{0})_{(0)}$, $\done \rho = -\done T^0_0$ are the unperturbed and first order stress tensor
respectively.  It is also conventional to define the comoving curvature perturbation at first order
\[
  \sR^\subo = \phi^\subo + \sH \frac{\done \varphi}{\varphi'_0}
\]
which, on large scales, is related to $\zeta^\subo$ as
\[
  \sR^\subo + \zeta^\subo \cong 0.
\]

The definition of the first order curvature perturbation is generally agreed upon in the literature (up to 
the sign of $\zeta^\subo$).  At second order, however, there are several definitions of the curvature 
perturbation in the literature (see \cite{lythNG} for a comprehensive discussion).  The definition we adopt 
follows \cite{Malik} and generalizes the definition of Malik and Wands \cite{MW} (valid
on large scales)
\beqa
  &-& \zeta^\subt \,
  = \, \psi^\subt + \frac{\sH}{\rho'_0} \dtwo\rho 
      - 2\frac{\sH}{(\rho'_0)^2}\done \rho \, \done \rho'  \\
  &-& 2 \frac{\done \rho}{\rho'_0}\left(\psi'^\subo +2\sH \psi^\subo \right) 
  + \frac{(\done \rho)^2}{(\sH \rho'_0)^2}
\left(\frac{\rho''_0}{\sH \rho'_0} - \frac{\sH'}{\sH^2} - 2
\right)\nonumber
\eeqa
to multiple scalar fields.  The definition of the curvature perturbation adopted in \cite{EV} generalizes 
\cite{Acquaviva} to two scalar fields.  However, it has been shown that this definition applies only during 
inflation \cite{Vernizzi,lythNG} since it is conserved only in the slow roll limit.

Because $\sigma_0=0$ the inflationary trajectory is straight the case of hybrid inflation and $\zeta^\subo$
is conserved on large scales \cite{BellidoWands}-\cite{Vernizzi}.  However, this is not the case at second 
order \cite{lythNG} and one expects that $\zeta^\subt$ will be amplified due to the 
tachyonic instability.  This amplification of $\zeta^\subt$ will continue until 
$N = N_\star$ at which
point the backreaction sets in and stops inflation.  For $N > N_\star$ the inflaton is no
longer dynamical since all of its energy has been converted into tachyon fluctuations.  Thus
for $N > N_\star$ the large scale curvature perturbation is conserved at all orders in 
perturbation theory since only one field (the tachyon) is dynamical and there are no 
non-adiabatic pressures.

In \cite{Malik} the second order large scale curvature perturbation is written in terms of the first and second order
Sasaki-Mukhanov variables as
\begin{eqnarray}
 \zeta^\subt &=& \frac{1}{3-\epsilon} \frac{1}{(\varphi'_0)^2}\left[ \varphi'_0 Q'^\subt_{\varphi} 
                 + a^2 \frac{\partial V}{\partial \varphi} Q^\subt_{\varphi} \right] \nonumber \\
             &+& \frac{1}{3-\epsilon} \frac{1}{(\varphi'_0)^2}\left[ \left(Q'^\subo_{\sigma}\right)^2 
                 + a^2 m_{\sigma}^2 \left(Q^\subo_{\sigma}\right)^2 \right] \nonumber \\
             &+& \frac{1}{3-\epsilon} \frac{1}{(\varphi'_0)^2}\left[ \left(Q'^\subo_{\varphi}\right)^2 
                 + a^2 m_{\varphi}^2 \left(Q^\subo_{\varphi} \right)^2 \right] \nonumber \\
             &+& 4(3+\epsilon-\eta)\left(\frac{3-2\epsilon}{3-\epsilon}\right) 
                 \left( \frac{\sH}{\varphi'_0} Q^\subo_{\varphi}\right)^2 \nonumber \\
             &+& (-10 + 2\epsilon + 2\eta) \left(  \frac{\sH}{\varphi'_0} Q^\subo_{\varphi}\right)^2
\label{zeta2}
\end{eqnarray}
where the first order Sasaki-Mukhanov variables
\footnote{Notice that $Q^\subo_{\varphi}$ is related to the variable $V^\subo$ discussed in subsection 
\ref{firstordersection} by $V^\subo = a \, Q^\subo_{\varphi}$.}
are
\begin{eqnarray}
  Q^\subo_{\varphi} &=& \frac{\varphi'_0}{\sH} \sR^\subo \label{Q1varphi} \\
  Q^\subo_{\sigma} &=& \done \sigma \label{Q1sigma}
\end{eqnarray}
and the second order Sasaki-Mukhanov variable is
\begin{eqnarray}
  Q^\subt_{\varphi} &=& \dtwo \varphi + \frac{\varphi'_0}{\sH}\psi^\subt \nonumber \\
                    &+& (2 + 2\epsilon -\eta) \frac{\varphi'_0}{\sH} \left(\phi^\subo\right)^2 \nonumber \\
                    &+& 2\frac{\varphi'_0}{\sH^2}\phi^\subo\phi'^\subo 
                        + \frac{2}{\sH}\phi^\subo\done\varphi'.
\label{Q2varphi}
\end{eqnarray}
In writing (\ref{zeta2}-\ref{Q2varphi}) we have restricted ourselves to hybrid
inflation, made use of the background equations and inserted 
$\epsilon,\eta$ (but not assuming slow roll).  

The last four lines of (\ref{zeta2}) are relatively simple to evaluate in the large scale and slow roll
limit.  On the other hand, the first line of (\ref{zeta2}) is somewhat more complicated and its evaluation 
requires the second order Einstein equations.  Thus we focus here on the following contribution to 
$\zeta^\subt$
\begin{equation}
\label{zeta2part}
  \zeta^\subt \ni \frac{1}{3-\epsilon}\frac{1}{(\varphi'_0)^2}\left[  \varphi'_0 Q'^\subt_{\varphi} 
                 + a^2 \frac{\partial V}{\partial \varphi} Q^\subt_{\varphi} \right].
\end{equation}
From the definition of $Q^\subt_{\varphi}$ it is clear that this contains a contribution of the form
$\varphi'_0 \dtwo \varphi' + a^2 \partial V / \partial \varphi \, \dtwo \varphi$ which also appears in the 
second order $(0,0)$ Einstein equation (\ref{00}).  We therefore use (\ref{00}) to eliminate $\dtwo \varphi$
from (\ref{zeta2part}).  We also eliminate $\psi^\subt$ in favour of $\phi^\subt$ using (\ref{phipsi}).
The result is that (\ref{zeta2part}) takes the form
\begin{eqnarray}
  \zeta^\subt &\ni& -\frac{\phi'^\subt}{\epsilon \sH}  - \left( \frac{1}{\epsilon} + 1 \right) \phi^\subt
                        + \frac{1}{3-\epsilon}\frac{\Lap \phi^\subt}{\epsilon \sH^2} \nonumber \\
              &+& \frac{1}{\epsilon \sH} \Linv \gamma' + \Linv \gamma 
                  - \frac{1}{3-\epsilon} \frac{1}{\epsilon \sH^2} \gamma \nonumber \\
              &+& \frac{1}{3-\epsilon} \frac{\Upsilon_1}{\epsilon \sH^2} 
                  + \frac{1}{3-\epsilon}\left(\frac{\Upsilon'_4}{\varphi'_0} 
                  + \frac{a^2}{(\varphi'_0)^2} \frac{\partial V}{\partial \varphi} \Upsilon_4 \right)
\label{zeta2firstline}
\end{eqnarray}
where the quantities $\Upsilon_1,\gamma$ are constructed entirely from first order fluctuations and are
defined explicitly in appendix B in equations (\ref{Upsilon1varphi}, \ref{Upsilon1sigma},
\ref{Upsilon2varphi}, \ref{Upsilon2sigma}, \ref{Upsilon3varphi}, \ref{Upsilon3sigma}).  The 
quantity $\Upsilon_4$ is also contructed from first order fluctuations
and is defined as the last two lines of (\ref{Q2varphi}).  
That is,
\begin{eqnarray}
  \Upsilon_4 &=& (2 + 2\epsilon -\eta) \frac{\varphi'_0}{\sH} \left(\phi^\subo\right)^2 \nonumber \\
             &+& 2\frac{\varphi'_0}{\sH^2}\phi^\subo\phi'^\subo + \frac{2}{\sH}\phi^\subo\done\varphi'.
\label{Upsilon4}
\end{eqnarray}

From (\ref{zeta2firstline}) we see that the dependence of $\zeta^\subt$ on the second order fluctuations
comes from the combination on the first line of (\ref{zeta2firstline}) which can be computed in terms of 
the source using the Green function (\ref{green}).

We split $\zeta^\subt$ into contributions coming from the inflaton and the tachyon as
\[
  \zeta^\subt = \zeta^\subt_{\varphi} + \zeta^\subt_{\sigma}
\]
and study each piece separately.  This splitting is
different from the one discussed in \cite{Malik}, where
the curvature perturbation is defined for each fluid in such a way
that the total curvature perturbation is a weighted sum of the
individual contributions.  Instead, we simply divide $\zeta^\subt$
into terms which depend respectively 
on the tachyon and inflaton fluctuations, $\done \sigma$  and 
$\done \varphi$, 
$\phi^\subo$,  which is only possible because
$\sigma_0=0$.  

The tachyon part of the curvature perturbation $\zeta^\subt_{\sigma}$ gets 
contributions from the first and second line of (\ref{zeta2}), both explicitly
through $Q^\subo_{\sigma}$ and implicitly through $\dtwo\varphi,\psi^\subt$.
The inflaton part of the curvature 
perturbation $\zeta^\subt_{\varphi}$ contains contributions from the
last three lines of (\ref{zeta2}) as  well as from the first line of
(\ref{zeta2}), both implicitly through $\dtwo\varphi,\ \psi^\subt$ and
explicitly through the definition of $Q^\subt_{\varphi}$.

The inflaton part of the curvature perturvation, $\zeta^\subt_{\varphi}$ coincides
with the $\zeta^\subt$of single field inflation and has been derived previously
\cite{Maldacena,SeeryLidsey,Calcagni}.  We have considered the construction of
$\zeta^\subt_{\varphi}$ using our formalism and these results are presented in 
appendix C.

\subsection{The Tachyon Curvature Perturbation}
\label{tachyonPart}

We now consider the tachyon contribution to the second order curvature
perturbation $\zeta^\subt_{\sigma}$.  It is
sourced by 
$J^{\sigma}$ which, in position space, takes the form (see equation (\ref{Tsource}))
\begin{eqnarray}
  J^\sigma(\tau,\vec{x}) &=& a^2 \kappa^2 m_{\sigma}^2 \left( \done \sigma \right)^2
                               - 2 \kappa^2 \left( \done \sigma' \right)^2 \nonumber \\
  &+& 2 \kappa^2 \sH (1+\eta-\epsilon) \Linv \partial_i \left( \done \sigma' \partial^i \done \sigma \right)
  \nonumber\\ &+& 4 \kappa^2 \Linv \partial_{\tau} \partial_i \left( \done \sigma' \partial^i \done \sigma \right) 
  \nonumber \\
  &-& \sH (1 + 2\epsilon - 2\eta) \Linv \gamma'_{\sigma} + \Linv \gamma''_{\sigma}.
  \label{tachyonSource}
\end{eqnarray}
The quantity $\gamma_{\sigma}$ can be written in the form (see equation (66) of \cite{EV},
or equuivalently (\ref{gamma_sigma2}))
\begin{eqnarray*}
  \gamma_{\sigma} &=& -\kappa^2 \Linv 
  \left[ 3 \partial_i \left( \Lap \done \sigma \partial^i \done \sigma
\right)\right.\nonumber\\
  &+& \left.
         \frac{1}{2} \Lap \left( \partial_i \done \sigma \partial^i \done \sigma \right) \right] \nonumber \\
               &=& -3 \kappa^2 \Linv \partial_i \left( \Lap \done \sigma \partial^i \done \sigma \right) \\
                &-& \frac{\kappa^2}{2} \left( \partial_i \done \sigma \partial^i \done \sigma \right).
\label{gamma_sigma}
\end{eqnarray*}
Notice that the terms in the first line of the (\ref{tachyonSource}) are local, the terms in the second
and third line are non-local (containing an inverse laplacian $\Linv$) and the fourth line contains terms
which are both local and doubly non-local (containing $\Linvd$).  The Fourier transforms of the source terms
are computed in appendix D wherein we also discuss our conventions for the inverse laplacian operators.

In the following, we will need the Fourier transform of terms like
$\Linv \gamma_{\sigma}$,
\begin{eqnarray*}
  \mathcal{F}\left[\Linv \gamma_{\sigma}\right] 
  &=& 
  -3\frac{\kappa^2}{k^4} \int \frac{d^3k'}{(2\pi)^{3/2}}
  k'^2 k\cdot (k-k') \done\tilde{\sigma}_{k'} \done\tilde{\sigma}_{k-k'} \\
  &-& 
  \frac{\kappa^2}{2k^2}\int \frac{d^3k'}{(2\pi)^{3/2}} 
  k'\cdot (k-k') \done\tilde{\sigma}_{k'} \done\tilde{\sigma}_{k-k'}
\end{eqnarray*}
This expression is 
operator-valued and can be written in terms of annihilation/creation operators
and mode functions as 
$\done\tilde{\sigma}_k = a_{k} \xi_k(t) + a_{-k}^{\dagger} \xi_{-k}(t)$ (see appendix D
for more details).  In the Fourier transformed expression for $\Linv \gamma_{\sigma}$, the scale 
dependence of the mode functions $\done \sigma_k$ is integrated over so that $\Linv \gamma_{\sigma}$ gets 
contributions from the tachyon fluctuations on all scales. The large scale limit of terms
like $\Linv\gamma_\sigma$ is not transparent and therefore we do not neglect any terms in
the tachyon source which contain inverse Laplacians.

We now compute $\zeta^\subt_{\sigma}$.  
The tachyon contribution to $\zeta^\subt$ comes about entirely through the implicit dependence of 
$\dtwo \varphi$ and $\psi^\subt$ on $\done \sigma$ in the first line of (\ref{zeta2}) and the explicit
dependence on the second line of (\ref{zeta2}).   Our focus is on the leading order contribution to
$\zeta^\subt$ in the slow roll and large scale limit.  If we work only to leading order in the slow roll
parameters it is sufficient to use the Green function (\ref{green}) in the limit $\epsilon=\eta=0$ and 
keep only the terms in the tachyon source (\ref{tachyonSource}) which
are not slow-roll suppressed.  
However, to consistently compute $\zeta^\subt_{\sigma}$ we must keep the next-to-leading order terms in
the small $(k/\sH)^2$ expansion of the Green function.  To see this notice that powers of
$k^2$ cancel inverse Laplacians in the source:
\[
  k^2 J^{\sigma}_k = -\gamma''_{\sigma,k} + \sH \gamma'_{\sigma,k} + \cdots
\]
where $\cdots$ denotes gradient terms which are small on large scales.
In appendix B it is shown that 
$\gamma_{\sigma}$ can be written on large scales as (\ref{gamma_sigma1})
\begin{eqnarray*}
  \gamma_{\sigma} &\cong& \frac{3\kappa^2}{2}\left[ \left(\done\sigma'\right)^2 
                            - a^2m^2_{\sigma}\left(\done\sigma\right)^2\right] \\
            &-& 3 \kappa^2\Linv\partial_\tau\partial_i\left(\done\sigma'\partial^i\done\sigma\right) \\
            &-&6\sH\kappa^2\Linv\partial_i\left(\done\sigma'\partial^i\done\sigma\right).
\end{eqnarray*}
Thus  $k^2 J^{\sigma}_k/\sH^2$ contains terms which are of the same form as those which appear
in $J^{\sigma}_k$ (\ref{tachyonSource}) and hence these terms must be included to consistently study 
$\zeta^\subt_{\sigma}$ on large scales.  

The curvature perturbation $\zeta^\subt$ depends on $\phi^\subt$ through the combination 
$-\phi'^\subt/\epsilon\sH - \phi^\subt/\epsilon + \Lap\phi^\subt/3\epsilon\sH^2$ at leading order in
slow roll parameters (\ref{zeta2firstline}).  Using the Green function for $\epsilon=\eta=0$ expanded up to 
order $k^2$ we find that
\begin{eqnarray}
  &-& \frac{\phi'^\subt_k}{\epsilon\sH} - \frac{\phi^\subt_k}{\epsilon} 
  - \frac{k^2\phi^\subt_k}{3\epsilon\sH^2} = 
  \nonumber \\
  &&\frac{1}{\epsilon}\int_{\tau_i}^0d\tau' \Theta(\tau-\tau')J_k^{\sigma}(\tau') \nonumber \\
  &\times& \left[
  \tau' + \left(-\frac{1}{6}\tau'^3 + \frac{5}{6}\tau^2\tau' - \frac{2}{3}\tau^3\right) k^2   \right] 
  \label{subleadinggreen}.
\end{eqnarray}

Integrating (\ref{subleadinggreen}) by parts\footnote{This integration by parts will 
give rise to terms which are  evaluated between the initial and final time
of the form $\left[\cdots\right]_{\tau'=-1/a_iH}^{\tau}$.   Since we
are interested in the preheating phase during which the fluctuations
are amplified exponentially we can safely drop the contribution at
$\tau'=-1/a_iH$ relative to the contribution at $\tau'=\tau$.} and
plugging the result into (\ref{zeta2}) and (\ref{zeta2firstline}) we
find that on large scales

\begin{eqnarray}
  \zeta^\subt_{\sigma} &\cong& \frac{\kappa^2}{\epsilon}\int_{-1/a_iH}^{\tau}d\tau' \left[
                               \frac{\left(\done\sigma'\right)^2}{\sH(\tau')} \right. \nonumber 
\\ &-& \left. \frac{\sH(\tau')^2}{\sH(\tau)^3}\left( \left(\done\sigma'\right)^2
- a^2 m_{\sigma}^2\left(\done\sigma\right)^2\right) \right]
\label{final}
\end{eqnarray}
where the tachyon fluctuations $\done \sigma$ are functions of the
integration variable $\tau'$.  The corrections to (\ref{final}) are
either subleading in the slow roll expansion or are total gradients
which can be neglected on large scales.   Equation (\ref{final}) is the main
result of this section.
The interested reader may find a more detailed discussion
of our calculation of $\zeta^\subt_\sigma$ in appendix E.

It is interesting to contrast the simplicity of this result with
other previous computations of ostensibly the same quantity
\cite{EV}.  Part of the difference is due to our different
definition of the second order curvature perturbation, but one must
also take considerable care  in keeping all terms which
are of the same order in slow roll parameters and powers of $k^2$
in order to obtain the delicate cancellations that collapsed the sum of
many terms down to this compact form.  Especially notable is the 
absence of nonlocal terms (containing inverse Laplacians) in the
final result, which were quite prevalent in the intermediate steps.
It is an important consistency check that such terms disappear in the
end, since they do not respect causality.  For example, a term
of the form $\Linv \partial_i\done\sigma\partial^i\done\sigma$
gives rise to an acausal response to a source which is localized in
time and space.  Consider a source of the form $\done\sigma \sim
e^{-\vec x^2/a^2}$ which is turned on at some instant in time.
Then $\Linv \partial_i\done\sigma\partial^i\done\sigma \sim 1/|\vec
x|$ instantaneously, at large distances, instead of being
exponentially small.  This would be a clear violation of causality 
and is physically inadmissible.

To clarify, we have shown that at leading order in slow roll parameters the nonlocal
contributions to $\zeta_\sigma^\subt$ cancel.  We have not checked that such terms
cancel at higher order in the slow roll expansion, though we believe that they do.
This cancellation was not observed in previous studies because the subleading (in $k^2$)
corrections to the Green function were not included and thus the large scale expansion
was inconsistent.

A comment is in order concerning the long wavelength approximation
which we have used in deriving (\ref{final}).  In writing the
expression for $k^2 J_k^{\sigma}$ we have dropped terms which are
total gradients  even though such terms will be integrated over time
in computing $\zeta^\subt_\sigma$, due to the time integral in
(\ref{particular}).  Strictly speaking, one should only apply the
large scale limit $k\tau \ll 1$ \emph{after} the time integral has
been performed.  The reason for this is that the time integral
extends from when the modes are well within the horizon, where they
oscillate, to when the modes are outside the horizon, including
horizon crossing.  Mode-mode coupling near the epoch of horizon
crossing can contribute momentum-dependent terms which are not
suppressed on large scales \cite{Maldacena,BBU}.   However in our
application,  we are justified in dropping  the contributions which
are generated near horizon crossing since they are exponentially
suppressed compared those which are produced during the
preheating phase.

\subsection{Time-integrated tachyon perturbation}
\label{sec_timeint}
We can make our main result (\ref{final}) more explicit 
by substituting in the solutions (\ref{quantum},\ref{mode}).
The result is simplified by taking the Fourier transform of
$\zeta^{(2)}$ and evaluating it at vanishing external wave number,
and at the final time corresponding to the end of inflation, $N=N_*$:
\beqa
\label{timeint}
	\zeta^{(2)}_{k=0}(N_*) &=& {\kappa^2\over\epsilon}
	\int {d^3 p \over(2\pi)^{3/2}} 
	\left(\,a_p\, b_p\, a^\dagger_p\, b_{-p}\,+ \hbox{perms}\right)
\nonumber\\&& 
\times\int_{{\rm max}(N_p,N_i)}^{N_*} f(c,N,N_*)\, dN 
\eeqa
where $N_i$ denotes the value of $N$ at the beginning of inflation,
$f(c,N,N_*)$ is given by
\beqa
\label{feqn}
 && f(c,N,N_{\ast}) = 
{e^{-3N + \frac{9}{2c}z^{3/2}}
\left(1 + |z|\right)^{-1/2}}\times\qquad\nonumber\\
 && \qquad\left[ \frac{9}{4}\left( 1-e^{3(N-N_{\ast})}\right)
     \left|\sqrt{z} - 1 - \frac{2 c\,\, {\rm sign}(z)}{27(1+|z|)} 
	\right|^2 - \right. \nonumber\\
&& \left. \qquad \phantom{\left(\frac94\right)^2_2} 
c N e^{3(N-N_{\ast})} \right]
\eeqa
where $z\equiv (1+\frac49cN)$, 
and ``perms'' in (\ref{timeint}) indicates the three other combinations of 
$a_p\, b_p$ and $a^\dagger_p\, b_{-p}$.

\begin{figure}[htbp]
\bigskip \centerline{\epsfxsize=0.5\textwidth\epsfbox{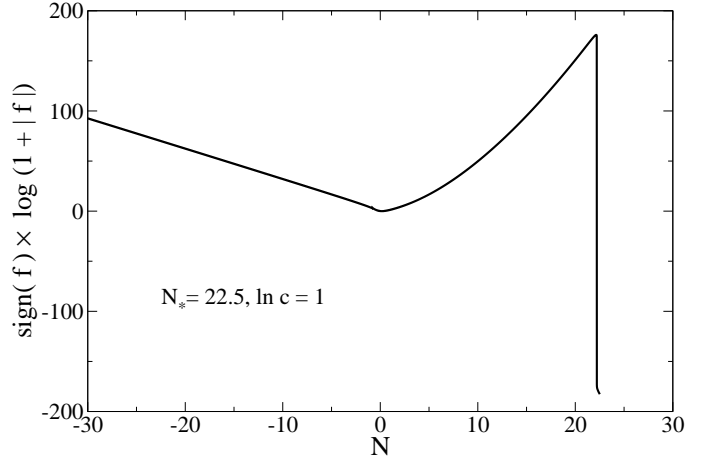}}
%\begin{verse}
%\vskip-0.25cm
\caption{$({\rm sign}(f)\ln(1+|f|)$ versus
$N$, showing behavior of the the function $f$ defined in (\ref{feqn}),
 for $\ln c = 1$, $N_*=22.5$}
\label{figxx}
\end{figure}

For illustration, we show the behavior of the function 
$f(c,N,N_{\ast})$ for sample parameter values $N_*=22.5$ and $\ln
c=1$.  Figure \ref{figxx} plots $({\rm sign}(f)\ln(1+|f|)$ as a
function of $N$.  The function is exponentially peaked at the initial
value $N=N_i$, and at the final value $N=N_*$.  Moreover it always 
becomes negative at $N_*$ because the negative mass squared term in 
(\ref{feqn}) comes to dominate.  Although it is not obvious in the
figure, the negative value is orders of magnitude larger than the
positive maximum just preceding it, so the negative extremum dominates
in the integral in (\ref{feqn}).  Whether the extremum at $N_*$ or
$N_i$ dominates overall depends on how $|3N_i|$ compares to $9/(2c)
z^{3/2}$ evaluated at $N_*$.  

Because of the exponential growth of $f$ at its extrema, it is a very
good approximation to the integral to write $f=e^g$ and expand $g =
g_m + g'_m(N-N_m)$ in the vicinity of the maximum value, whether it is
at $N_i$ or $N_*$.  Since the integral is so strongly peaked near the
extremum, there is an exponentially small error in extending
the range of integration to the half-line. In this way one obtains
\beq
\label{intapp}
	\int dN\, f \cong {e^{g_m}\over |g'_m|}
\eeq
We will use this approximation below to numerically evaluate the
integral.

\section{Bispectrum  and spectrum of second order metric perturbation}
\label{V}
\label{sect:bi}

Here we calculate the leading contribution to the three-point function
(bispectrum) of the second-order curvature perturbation due to the tachyon,
\beq
\label{bispectrum}
\left\langle \zeta^{(2)}_{k_1}\, \zeta^{(2)}_{k_2}\,
 \zeta^{(2)}_{k_3} \right\rangle 
\equiv (2\pi)^{-3/2}\, \delta^{(3)}(\vec k_1 + \vec k_2 + \vec k_3)\,
B(\vec k_1, \vec k_2 ,\vec k_3)
\eeq
It is understood that only the connected part of the correlation
function is computed, which is equivalent to subtracting the expectation
value of $\zeta^{(2)}$ from the quantum operator.  
%Since there is no
%momentum dependence in the result (\ref{int2}) for $\zeta^{(2)}$,
This three-point function is straightforward to compute, using the
free-field two-point functions
\beq
\left\langle \done\tilde{\sigma}_{p_i} \done\tilde{\sigma}_{q_i} 
\right\rangle = 
\xi_{p_i}\xi_{q_i}\,\delta^{(3)}(p_i + q_i) 
\eeq

\begin{figure}[htbp]
\bigskip \centerline{\epsfxsize=0.5\textwidth\epsfbox{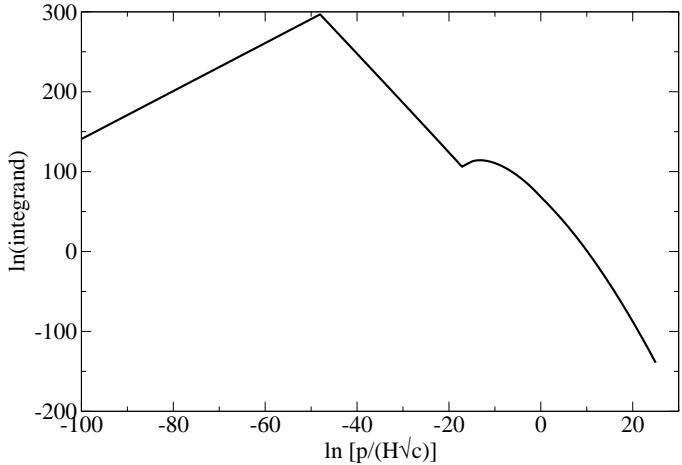}}
%\begin{verse}
%\vskip-0.25cm
\caption{Log of the integrand of the $p$ integral in eq.\ 
(\ref{Beq}),
for a case where the maximum occurs near $\ln(p/\sqrt{c} H)= N_i$
.}
\label{figww}
\end{figure}

\begin{figure}[htbp]
\bigskip \centerline{\epsfxsize=0.5\textwidth\epsfbox{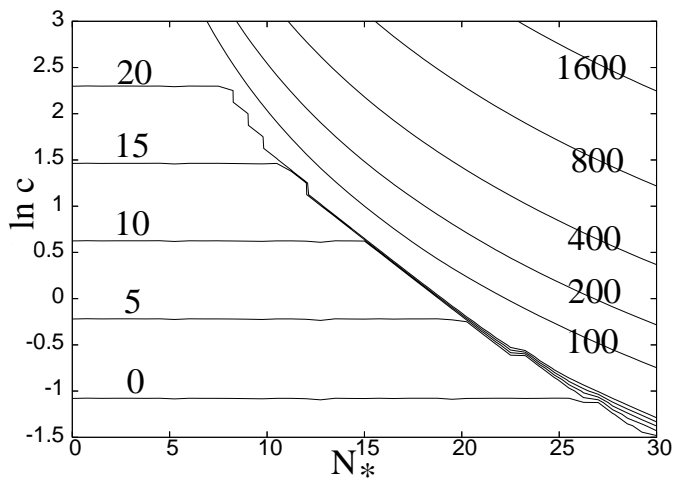}}
%\begin{verse}
%\vskip-0.25cm
\caption{Contours of $\ln|\tilde B|$, defined in (\ref{tb}),
in the plane of $N_*$ and $\ln c$.  $N_i = -30$ in this example.}
\label{figyy}
\end{figure}

Carrying out the contractions of pairs of fields which contribute
to the connected part of the bispectrum, one finds eight terms,
which in the limit of vanishing external wave-numbers, are all equal.
The result is
\beq
\label{Beq}
	B = 8{\kappa^6\over\epsilon^3}\int {d^{\,3}p\over (2\pi)^3} |b_p|^6\left
[\int_{{\rm max}(N_p,N_i)}^{N_*}\!\!\!\!\!\!\!\! dN\, f(c,N,N_*)\right]^3
\eeq
The integrand of the $p$-integral is exponentially strongly peaked,
either near $\ln (p/\sqrt{c} H)\cong N_i$ if $f$ has its global maximum
near $N=N_i$, or else near $\ln (p/\sqrt{c} H)\cong 0$  if $f$ is
dominated by its behavior near $N_*$.  The logarithm of the integrand
for a typical case is shown in figure \ref{figww}.
If the cusp-like peak is dominating,  we can use the same
approximation as in (\ref{intapp}) to evaluate the $p$ integral,
on both sides of the maximum.  If the other local maximum at larger
values of $p$ is the global maximum, as sometimes happens, then we
should treat the integral as a gaussian, since the derivative of
the integrand vanishes at the maximum.  In this case we similarly
write the integrand of the $p$ integral as $f=e^g$ and expand $g =
g_m + \frac12 g''_m(p-p_m)$, where $g''_m<0$; then (\ref{intapp})
is replaced by
\beq
\label{intapp2}
	\int dN\, f \cong \sqrt{2\pi\over |g''_m|}e^{g_m}
\eeq		
We have carried out the evaluations
numerically over a range of values of $\ln c$ and
$N_*$.  In figure 
(\ref{figyy}) we plot contours of $\ln|\tilde B|$,
where $\tilde B$ is defined by
\beq
\label{tb}
	B = {e^{-6N_i}\over 2\pi^2} \left({\kappa^2\over c\epsilon}\right)^3
	\tilde B
\eeq

Although we evaluated $B$ at vanishing external wave numbers, 
$k_i = 0$,  the result is equivalent to evaluating  $B$ at 
very small wave numbers closer to horizon crossing, 
$k_i \cong H e^{N_i}$.
This is not immediately obvious, but one can see that at very small
$p$, the $p$ integral in (\ref{Beq}) behaves like
$\int d^3 p/p^9$, which in the presence of nonvanishing external
wave numbers would be $\int d^3 p\,p^{-3} (p-k)^{-6}\sim k^{-6}$.
(This claim is explicitly verified in appendix F.)  
However, (\ref{Beq}) does not diverge as $k\to 0$; the $p$ integral
is cut off near $He^{N_i}$ because of the lower limit max$(N_i,N_p)$
in the integral over $N$.  If the lower limit were simply $N_p$
instead of max$(N_i,N_p)$, the value of the $N$ integral near its
lower limit would go like $e^{3N_p} \sim 1/p^3$ instead of $e^{N_i}$.
Thus the factor of $e^{6N_i}$ in (\ref{tb}) is a reflection of the
fact that the bispectrum (at least the part of it which is due to the
low-$p$ part of the integral) is scale-invariant, $B\sim 1/k^6$,
for $k > He^{N_i}$, but the infrared divergence is cut off for
$k < He^{N_i}$, and the spectrum remains flat as $k\to 0$.  
 
A consequence of the above discussion is that the bispectrum is 
scale-invariant only in the part of parameter space where the 
low-$N$ part of the $N$ integral dominates, namely the lower
left-hand region of fig.\ (\ref{figyy}).  The upper right-hand
region of fig.\ (\ref{figyy}) gets its dominant contributions from
the large-$p$ part of the $p$ integral, which is not infrared
sensitive.  The bispectrum really is flat as a function of $k$, for
small $k$, in this region.  Furthermore, $B$ is negative in this
region, whereas it is positive in the other region.  This means there
is a very narrow, fine-tuned curve along which $B$ vanishes, seen
as the borderline between the two different behaviors.  This is a
region which will always be allowed as long as no nongaussianity is
observed in the CMB, but since it is a set of measure zero in the
parameter space, we will not take it into account in deriving limits
below.

\begin{figure}[htbp]
\bigskip \centerline{\epsfxsize=0.5\textwidth\epsfbox{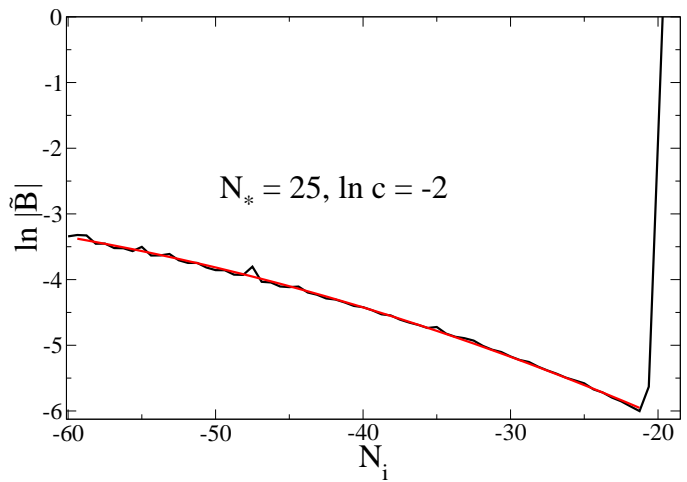}}
%\begin{verse}
%\vskip-0.25cm
\caption{Dependence of $\ln|\tilde B|$ on $N_i$ for 
$N_*=25$ and $\ln c=-2$. The smooth curve is the numerical
fit $\ln|\tilde B| = -8.3 - 0.13N_i - 0.0007N_i^2$.}
\label{figzz}
\end{figure}

The boundary between the two regions in fig.\ (\ref{figyy}) is
described analytically by the relation 
\beq
	3|N_i| = {9\over 2c}z_*^{3/2}
\eeq
with $z_* = 1 + \frac49 c N_*$.  This where the two terms in the 
exponent of $f$ in (\ref{feqn}), which determinant the dominant
behavior of $f$, just balance.  In the region to the right, where
${9\over 2c}z_*^{3/2} > 3|N_i|$, which is also where $B$ is
independent of $k_i$, $B$ is also independent of $N_i$.  In the region
to the left, $B$ depends principally on $N_i$ through the factor
$e^{-6N_i}$ in (\ref{tb}).  However there is some mild extra
dependence on $N_i$, as shown in figure (\ref{figzz}).

To make contact with experimental constraints, we want to compare the 
predicted bispectrum with that of single-field inflation, where
nongaussianity is conventionally expressed via a nonlinearity
parameter $f_{NL}$, defined through
\beq
\label{standard}
	B(k_i) = -{6\over 5} f_{NL} \left( P_{\phi}(k_1)\,
	 P_{\phi}(k_2) +\hbox{\ permutations} \right)
\eeq 
where $P_{\phi}(k)$ is the usual inflationary power spectrum,
$P_{\phi}(k)^{1/2} \sim  10^{-5}\,(2\pi) / k^{3/2}$. If we assume that
all $k_i$ are the same, $k_i\cong k$, then
\beq
\label{feq}
	f_{NL} = - {5\over 18}(2\pi)^{-4} 10^{20}\,B(k)\, k^6
\eeq
We can thus convert our predicted bispectrum into an effective 
$f_{NL}$, for which the present experimental constraint is
roughly $|f_{NL}| < 100$.  To get the strongest constraint,  
$k$ should be evaluated at scales near horizon crossing in the
scale-invariant region of $B$, and at the smallest relevant scales
in the region where $B$ is $k$-independent.  Since $B(k)$ is evaluated
at the end of inflation, the horizon-crossing scale is $k\sim H
e^{-N_e}$.  On the other hand, current experimental constraints on 
nongaussianity involve temperature multipoles of the CMB going up to
$\ell = 265$ \cite{komatsu}, which represents scales roughly $e^5$ times smaller than
horizon-crossing.  So we should use $k=e^{-N_e+5}H$ in the region
where $B$ is constant at low $k$. It turns out that this is the region
of parameter space where the experimental limit is saturated, so it
is the relevant case.  

\begin{figure}[htbp]
\bigskip \centerline{\epsfxsize=0.5\textwidth\epsfbox{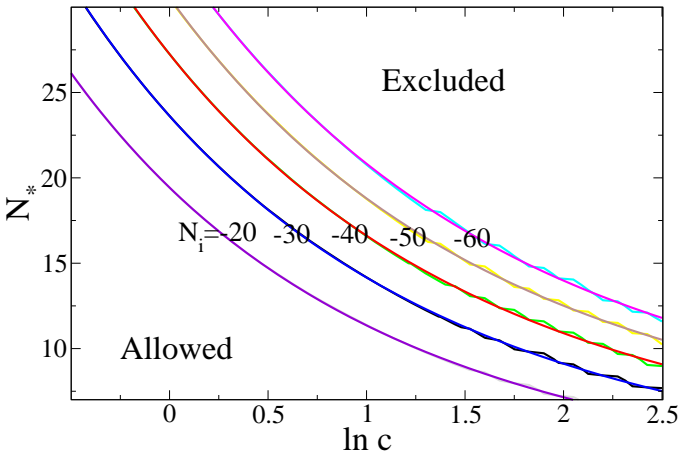}}
%\begin{verse}
%\vskip-0.25cm
\caption{Boundaries of the regions excluded by the nongaussianity
constraint $|f_{NL}|<100$, in the plane of $\ln c$ and $N_*$, for different values of $N_i$. }
\label{fnll}
\end{figure}

Using (\ref{tb}) in (\ref{feq}) and recalling that $N_e = |N_i|+N_*$
we find that 
\beq
	f_{NL} = -e^{17.5-6N_*}c^{-3}\tilde B
\eeq
where we used $H^2 = V/(3M_p^2)$ as well as the 
COBE normalization $V/(\epsilon M_p^4) = 6\times 10^{-7}$.  The
contours of $\tilde B$ can thereby be converted into contours of
$f_{NL}$, and demanding that $|f_{NL}|<100$ gives a constraint in 
the parameter space $c,N_*,N_i$.  
The limiting  curves
are shown for different values of $N_i$ in figure \ref{fnll}.  These
curves are quite insensitive to the actual value assumed for
$f_{NL}$, because the effect turns on exponentially fast.  The curves
for $f_{NL}=1$ are visually hard to distinguish from those shown
(for $f_{NL}=100$).

\subsection{Constraint from tachyon contribution to spectrum}

The analogous calculation can also be carried out for the tachyon
contribution to the two-point function of the curvature perturbation.
We show that it gives a stronger constraint than does nongaussianity.

\begin{figure}[htbp]
\bigskip \centerline{\epsfxsize=0.5\textwidth\epsfbox{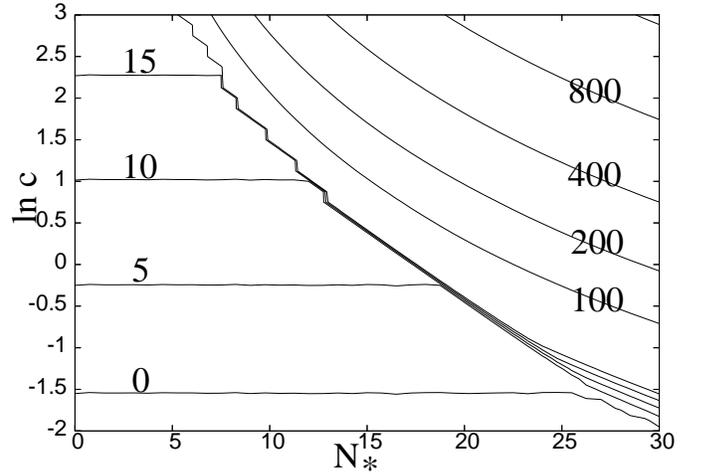}}
%\begin{verse}
%\vskip-0.25cm
\caption{Contours of $\ln|\tilde S|$ in the plane of $N_*$ and $\ln c$,
at $N_i=-30$. }
\label{fig2}
\end{figure}

Closely following the preceding calculation, it is straightforward to
show that, in the limit of vanishing external wave numbers, 
\beqa
\left\langle \zeta^{(2)}_{k_1}\, \zeta^{(2)}_{k_2}\,
\right\rangle_{\rm con} &=&
\delta(\vec k_1 + \vec k_2)\times  \nonumber\\
2{\kappa^4\over\epsilon^2}\int {d^{\,3}p\over (2\pi)^3} |b_p|^4
\!\!\!\!\!\!&&\!\!\!\!\!\left
[\int_{N_p}^{N_*} dN\, f(c,N,N_*)\right]^2\nonumber\\
\qquad\qquad	&\equiv& \delta(\vec k_1 + \vec k_2) \, S(k_i)
\label{spectrum}
\eeqa
In analogy to (\ref{tb}), we define\footnote{The powers of $H$ and 
${c}$
can be understood as follows:  $b_p\sim H^{-1/2}c^{-3/4}$,
whereas $p\sim \sqrt{c} H$.}
\beq
	S = {e^{-3N_i}\over 4\pi^2}\, {H\over c^{3/2}}\, \left({\kappa^2\over
\epsilon}\right)^2\tilde S
\label{ts}
\eeq
and we display contours of $\tilde S$ in fig.\ \ref{fig2}.
Analogously to the bispectrum, $S$ is approximately scale-invariant 
($\sim 1/k_i^3$) in the left-hand region, 
but independent of $k$ in the right-hand region.

The COBE normalization of the power spectrum implies that 
\beq
\label{ineq}
\left\langle \zeta^{(2)}_{k_1}\, \zeta^{(2)}_{k_2}\,
\right\rangle_{\rm con} \le {2\pi^2\over k^3}\, \delta({\vec k_1+\vec k_2})
\, {\cal P}_\zeta
\eeq
with ${\cal P}_\zeta^{1/2} \cong 10^{-5}$. Combining this with
(\ref{spectrum}) and (\ref{ts}), we get an upper bound
on the quantity
\beq
\label{fleq}
	f_L\equiv {10^{20}\over (2\pi)^{3}} \, {H\over c^{3/2}}\,
\left({\kappa^2\over
\epsilon}\right)^2\, \tilde S\, e^{-3N_i}\,k^3 < 1
\eeq
where the strongest constraint is obtained by taking the largest
$k$ values which are measured by the CMB.  We will 
conservatively take this to be $k= e^6 H e^{-N_e}$.  
Notice that the tachyon fluctuations have a spectral index of
$n=4$ (defined by $k^3 S \sim k^{n-1}$) which is consistent 
with the Traschen integral constraints \cite{traschen}.
Following the
same steps as we did for the bispectrum, this can be rewritten as
\beq
\label{fleq2}
	f_L = e^{30-3N_*}\, c^{-3/2}\, \tilde S \left\{
\begin{array}{cc} 1, & {9\over 2c}z_*^{3/2} > 3|N_i|\\
	e^{-18},& {9\over 2c}z_*^{3/2} < 3|N_i|\end{array}
\right\}
< 1
\eeq

\begin{figure}[htbp]
\bigskip \centerline{\epsfxsize=0.5\textwidth\epsfbox{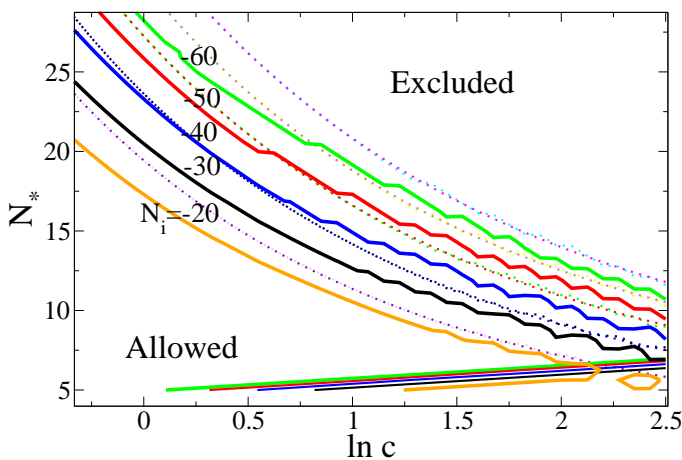}}
%\begin{verse}
%\vskip-0.25cm
\caption{Boundaries of the regions excluded by the spectral
constraint $|f_{L}|<1$, in the plane of $\ln c$ and $N_*$, 
for different values of $N_i$.  The corresponding constraints
from $f_{NL}$ (fig.\ \ref{fnll}) are shown as dotted lines.
}
\label{fll}
\end{figure}

One might consider being more conservative and imposing, say, $|f_{L}| < 0.01$ rather
than $|f_{L}| < 1$, as we have done.  Our exclusion plots are actually quite insensitive
to the value assumed for $f_{L}$, as was the case with $f_{NL}$, since the effect turns
on exponentially fast.

\section{Constraints on hybrid inflation model parameters}
\label{VI}
In the previous section we computed the effective nonlinearity
parameter $f_{NL}$ which characterizes the size of nongaussian
fluctuations produced by tachyonic preheating in the hybrid inflation
model under consideration.  Experimentally, it is currently
constrained as $|f_{NL}|\lsim 100$
\cite{komatsu}, which is expected to improve to the level
of $|f_{NL}| < 5$ from future experiments, like PLANCK
\cite{NGreview, KS}. Similarly, we defined
a parameter $f_L$ which characterizes the size of the second-order 
tachyonic contribution to the spectrum relative to the experimental
value, with the constraint $|f_L| < 1$.  These constraints were 
most easily determined in terms of derived parameters $c$ and $N_*$,
but here we want to recast them in terms of the fundamental parameters
of the hybrid inflation model, the dimensionless couplings
 $g$ and $\lambda$, and the VEV of the tachyon field, expressed 
in the dimensionless combination $v/M_p$.  We do not treat 
$m^2_\varphi$ as an independent parameter, since we used the COBE
normalization to eliminate it in (\ref{cobe}).

The only obstacle to working directly in the model
parameter space is the implicit dependence of $N_*$ on 
$(\lambda,g,v)$.  We have numerically inverted the relation
depicted in fig.\ (\ref{figNs}) to determine $N_*(\lambda,g,v)$.
It then becomes straightforward to scan the model parameter space,
recomputing $f_L$ and $f_{NL}$.  We display the constraints in the
plane of $\log_{10} g$ and $\log_{10}\lambda$, for a range of values
of $\log_{10}(v/M_p)$.  In performing this scan, one must keep track 
of whether various assumptions are respected.  These include
eqs.\ (\ref{lincon}) and (\ref{glb}), respectively 
corresponding to the assumption that $m^2_\sigma$ varies linearly with
the number of e-foldings, and that the vacuum energy density is
dominated by $\lambda v^4/4$ during inflation.  In addition to these,
we implement two others:  we assume that (1) the reheat temperature
exceeds 100 GeV, so that baryogenesis can occur at least during the
electroweak phase transition;  and (2) we limit ourselves to
considering values $c \ge 10^{-4}$, which entails a lower limit on
the coupling $g$ depending on $v$, eq.\ (\ref{vend}).
This is a numerical limitation,
due to the difficulty of evaluating the integral (\ref{vend}) for
small $c$.  However, analytic calculations can be done in this
region, which we have carried out in Appendix F, as
we describe below.

In the region where the condition (\ref{lincon}), $g v/M_p < 10^{-5}$, 
is not satisfied
our  analysis breaks down because the tachyon mass $m^2_{\sigma}$
varies exponentially rather than linearly with the number of
e-foldings.  However, we do not expect that any significant
nongaussianity will be generated in the region where $c$ is too large
because in this case we will have had $m^2_{\sigma} > H^2$ for a
significant number of e-foldings before the instability sets in and
the fluctations of the tachyon  will have a large exponential
suppression. However, to be conservative in our analysis we consider
the region where (\ref{lincon}) is not satisfied as allowed by the
experimental constraints on $f_{L}, f_{NL}$.  This is a rather small
sliver of parameter space, since the spectral index contstraint
(\ref{etacon}) requires that $g v/M_p < 5\times 10^{-5}$.

Once a set of values for $\lambda$, $g$ and $v/M_p$ are chosen
and $N_*$ has been calculated, one must still determine $N_i$
in order to calculate the parameters $f_L$ and $f_{NL}$.  We do
so by first computing the total number of e-foldings using the
standard result
\beq
        N_e = 62 - \ln\left({10^{16}{\rm\ GeV}\over V^{1/4}}\right)
	-\frac13 \ln\left({V^{1/4}\over \rho_{\rm r.h.}^{1/4}}\right)
\eeq
where $V \cong \lambda v^4/4$ is the energy density during inflation,
and $\rho_{\rm r.h.}$ is the energy density at reheating.  In the
following, we will ignore the gravitino bound 
$\rho_{\rm r.h.}^{1/4}\lsim 10^{10}$ GeV and assume instant
reheating, $\rho_{\rm r.h.} = V$.  The value of $N_i$ then follows
from $N_i = N_*-N_e$.  We have checked that incorporation of the
gravitino bound does not create a noticeable change in the excluded
regions.

\begin{figure}[htbp]
\bigskip \centerline{\epsfxsize=0.5\textwidth\epsfbox{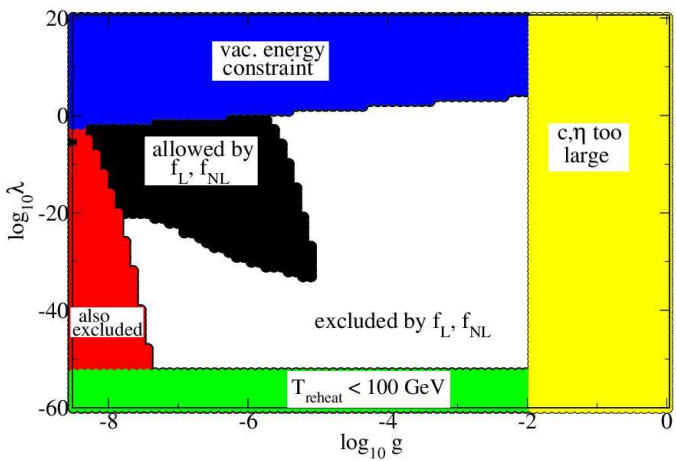}}
%\begin{verse}
%\vskip-0.25cm
\caption{Unshaded region in the plane of $\log_{10} g$ and $\log_{10}
\lambda$ is excluded by the tachyon contribution to the second order
curvature perturbation, for $v/M_p = 10^{-3}$.  Leftmost region
is also excluded by a separate analysis (Appendix F).    
}
\label{figcon}
\end{figure}

The resulting constraints are illustrated for the case where
$v/M_p = 10^{-3}$ in figure \ref{figcon}.  The unshaded region
is excluded by our calculation of $f_L$ and $f_{NL}$.  The shaded
region on the right falls outside our approximation (\ref{lincon})
or the spectral index constraint (\ref{etacon}),
that on top fails to satisfy (\ref{glb}), and
the region on the bottom has too low a reheat
temperature.  The region on the left has $c < 10^{-4}$ and we
were not able to treat it using the same numerical methods, but
we analyzed this region analytically in Appendix F.
The tachyon mass is small throughout inflation in this region, and
gives rise to a scale-invariant spectrum of nongaussian fluctuations,
unlike the excluded region shown in figure \ref{figcon}, which
corresponds to a scale-noninvariant contribution, with spectral
index $n=4$.  Nevertheless, the magnitude of nongaussianity
in this region labeled, ``also excluded,'' exceeds the experimental
bounds.

\begin{figure}[htbp]
\bigskip \centerline{\epsfxsize=0.5\textwidth\epsfbox{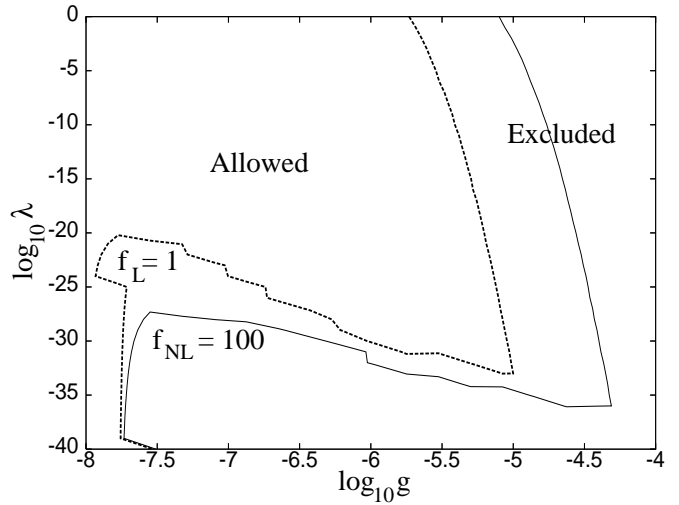}}
%\begin{verse}
%\vskip-0.25cm
\caption{Closeup of the borderline region in figure \ref{figcon},
showing that the spectral distortion  ($f_L<1$, dashed line) provides the stronger
constraint, relative to nongaussianity ($f_{NL}<100$, solid line).
}
\label{figcon2}
\end{figure}

As was indicated in the previous section, we get a stronger
constraint from the spectrum, whose distortions are parametrized by
$f_L$ (\ref{fleq}), than from nongaussianity, parameterized by
$f_{NL}$.  This is shown in figure \ref{figcon2}, which plots the
contours for the constraints $f_L<1$ and $f_{NL}<100$ again in the
plane of  $\log_{10} g$ and $\log_{10}\lambda$, and for 
$v/M_p = 10^{-3}$.

\begin{figure}[htbp]
\bigskip \centerline{\epsfxsize=0.5\textwidth\epsfbox{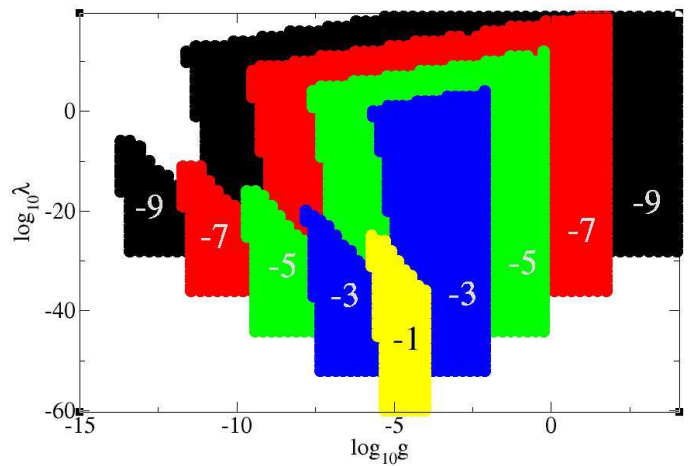}}
%\begin{verse}
%\vskip-0.25cm
\caption{Regions excluded by spectral distortion for 
$\log_{10}v/M_p = -1,-3,-5,-7,-9$, as labeled on the figure.
Note that for each $v$, the regions at smaller $g$ are also
excluded, but due to nongaussianity (Appendix F).}
\label{figcon3}
\end{figure}

The size and position of the excluded region is rather sensitive
to the value of the VEV, $v$.  We have varied $\log_{10}v/M_p$
between $-1$ and $-9$ to show how the constraints change with 
$v$ in figure \ref{figcon3}.  One sees that the width of the
excluded region for $\log_{10}g$ grows linearly with 
$\log_{10}M_p/v$, so the constraints are strongest at the smallest
values of $v$. 

\section{Implications for Brane-Antibrane Inflation}
\label{VII}

We now consider what implications our analysis has for the case of
brane-antibrane inflation \cite{DvaliTye}-\cite{multibrane}, which is
a string theoretic realization of hybrid inflation.  In this scenario
inflation is driven by the interactions  between a brane and
antibrane which are parallel to the three large dimensions and are 
initially separated in the extra dimensions of string theory.  The
brane-antibrane pair experiences an attractive force and move towards
each other with inflation ending when the branes collide and
annihilate.  The role of the inflaton $\varphi$ is played by  the
interbrane separation while the field $\sigma$ is the lightest
streched string mode between the brane and antibrane, which develops a
tachyonic mass when the separation between brane and antibrane
becomes of order the string scale.

The potential of brane-antibrane inflation is not ammenable to the form (\ref{pot}) for
several reasons:
\begin{itemize}
  \item The inflaton potential typically has the Coulombic form $V
\sim A/\varphi^n$, 
        rather than the simple form $V \sim m_{\varphi}^2\varphi^2$.
  \item The tachyonic field is complex, rather than real.
  \item The tachyon potential does not have its minimum at a finite value of the $\sigma$
        field.
  \item The tachyon field is described by a Dirac-Born-Infeld (DBI) action rather than
        a simple Klein-Gordon action as we have assumed in this paper.
\end{itemize}
Here we will assume that these differences do 
not significantly alter the analysis and
leave a more detailed investigation for the future.  
Specifically, this amounts to assuming that the behavior of
the two models will be roughly the same if we match the quadratic
parts of their actions, as well as the cross-coupling between the
inflaton and the tachyon.  Since the timescale for tachyon
condensation is given by $\sqrt{-m^{-2}_\sigma}$, one might hope that
the results are rather insensitive to the details of how the tachyonic
instability saturates at large field values.  

In string theory the open string tachyon $T$ between a D$p$-brane and antibrane, separated by
a distance $y$ is described by the action \cite{Sen}
\begin{eqnarray}
  S_{\mathrm{tac}} &=& -\int d^{p+1}x \, \sqrt{-g} \, \mathcal{L} \nonumber \\
  \mathcal{L} &=& V(T,y) \, \sqrt{1 + M_s^{-2} g^{\mu\nu}\partial_{\mu}T\partial_{\nu}T^{\dagger}}
  \label{DBI}
\end{eqnarray}
where the small $|T|$ expansion of the potential is
\begin{eqnarray}
  V(T,y) &=& \tau_p \left[ 1 + 
           \frac{1}{2} \left(\frac{(M_s y)^2}{(2\pi)^2} - \frac{1}{2}\right)|T|^2 \right. \\
         && \,\,\,\,\,\,\,\,\, + \left. \mathcal{O}(|T|^4) + \cdots 
	\phantom{\frac{(M_s y)^2}{(2\pi)^2}}
\!\!\!\!\!\!\!\!\!\!\!\!\!\!\!\!\!\!\!\right].
\label{smallT}
\end{eqnarray}
In the above we have neglected the gauge fields and transverse scalars on the branes.
The coefficient of the $|T|^4$ term in the potential is order unity and its exact 
numerical value is unimportant for our estimates since the potential $V(T,y=0)$ 
is minimized only at infinite $T$.  A potential of the form 
\[
  V(T,y=0) = \tau_p \, e^{-|T|^2/a^2}
\]
is typical.
In the above $M_s$ is the fundamental string mass and $\tau_p$ is the D$p$-brane tension.
In order to obtain order-of-magnitude estimates for the model parameters $\lambda,g,v$
of (\ref{pot}) it is sufficient to consider only the tachyon part of the
action (\ref{DBI}) and neglect the inflaton potential.
We restrict ourselves to inflation models driven by D$3$-branes since inflation driven
by higher dimensional branes have problems with overclosure of the universe by defect 
formation \cite{BBCS}.  The D$3$-brane tension is given by
\begin{equation}
  \tau_3 = \frac{M_s^4}{(2\pi)^3 g_s}
\end{equation}
where $g_s$ is the string coupling.

\subsection{The KKLMMT Model}

Brane inflation models in torodial compactifications are not realistic because the moduli
which describe the size and shape of the extra dimensions were assumed fixed without
specifying any mechanism for their stabilization.  Recently there has been significant
progress \cite{KKLMMT,realistic} in reconciling brane inflation with modulus stabilization 
using warped geometries with background fluxes for type IIB vacua \cite{vacua}.  For our 
purposes the most important feature of these compactifications is the presence of 
strongly-warped throats within the extra dimensions.  The compactification geometry within 
the throat is well approximated by
\begin{equation}
\label{AdS5}
  ds^2 = a(y)^2 g_{\mu\nu}dx^{\mu}dx^{\nu} + dy^2 + y^2 d\Omega_5^2
\end{equation}
where $y$ is the distance along the throat, $a(y) = e^{-ky}$ is the throat's warp factor
and $d\Omega_5^2$ is the metric on the base space of the corresponding conifold 
singularity of the underlying Calabi-Yau space \cite{base}.  In the subsequent analysis we
ignore the base space and treat the geometry as $\mathrm{AdS}_5$.  

A mobile D$3$-brane is placed near the UV end of the throat $y=0$ while an 
anti-D$3$-brane remains fixed at some location near the IR end of the throat 
$y = y_i > 0$.  The warp factor $a_i = a(y_i)$ can be easily made much less than
unity by a suitable choice of background fluxes and this large warping flattens the
inflaton potential.  This large warping also suppresses the tachyon potential 
(effectively reducing the brane tension to $a_i^4 \tau_3$).  At quadratic order, the tachyon action
takes the form
\begin{eqnarray}
  S &=& -\int d^4x \sqrt{-g}\left[ a_i^4 \tau_3 + 
        \frac{a_i^4\tau_3}{2}
        \left(\frac{M_s^2y^2}{(2\pi)^2} - \frac{1}{2}\right)|T|^2
\right.\nonumber \\
     &-& \left. \frac{a_i^{2}\tau_3}{2M_s^2}|\partial T|^2
         \right]
\label{smallTw}
\end{eqnarray}

The effective values of the couplings
can be found by rewriting the action (\ref{smallTw}) in
terms of the canonically normalized fields $\sigma = a_i\sqrt{\tau_3} T
/ M_s$ and $\varphi = \sqrt{\tau_3} y$ 
(see equations 3.6, 3.10 or C.1 in \cite{KKLMMT}), and then matching 
to the hybrid inflation potential (\ref{pot}).
This gives the correspondence
\begin{eqnarray*}
  v &=& \frac{a_i}{\pi^{3/2}}\frac{M_s}{\sqrt{g_s}} \\
  \lambda &=& \frac{\pi^3}{2} g_s \\
  g &=& a_i \sqrt{2\pi g_s}
\end{eqnarray*}

Before trying to make a general investigation of the parameter space, we 
will examine the fiducial values which were considered 
natural in \cite{KKLMMT}: $a_i = 2.5\times 10^{-4}$, $g_s=0.1$, $\tau_3/M_p^4 =
10^{-3}$.  The COBE normalization also fixes the inflationary energy scale to be
$\Lambda =1.3\times 10^{14}$ GeV in this model.  One finds that these values imply
\beq
\label{fid}
	{v\over M_p} = 10^{-4.2},\quad \lambda = 10^{0.2},\quad g = 10^{-3.7}
\eeq
By recomputing the excluded region in the $\lambda$-$g$ plane for this value of
$v/M_p$, we find this point to be solidly excluded. 

\begin{figure}[htbp]
\bigskip \centerline{\epsfxsize=0.5\textwidth\epsfbox{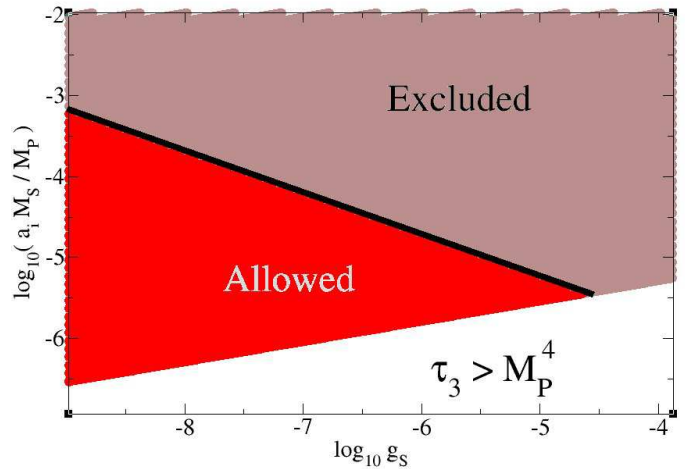}}
%\begin{verse}
%\vskip-0.25cm
\caption{Allowed and excluded values of the warped string scale and string coupling
in the KKLMMT model, due to tachyonic preheating.  The fiducial
point corresponding to (\ref{fid}) is at $(-1,-3.9)$, off the scale
of this graph, but in the excluded region.}
\label{figkklmmt}
\end{figure}

To get out of the excluded region, one should try to achieve smaller values of 
$g$ or larger values of $v/M_p$. One can show from the above relations that
\beq
	{v\over M_p} = \left({8\over \pi^6 g_s}\right)^{1/4} {\Lambda\over M_p}
\eeq
Since $\Lambda$ is fixed by the COBE normalization, lowering the string coupling is
the only way to increase $v/M_p$. Decreasing $g_s$ also has the desired effect of 
reducing $g$.  The other parameter we can adjust is the warp factor, but there
is not much leeway for decreasing $a_i$ because the density contrast in this model
is given by \cite{KKLMMT}
\beq
	\delta_H = C_1 N_e^{5/6}\left({\tau_3\over M_p^4}\right)^{1/3} a_i^{4/3}
\eeq
where $C_1$ is a constant, and $N_e\cong 60$.  (Lower values of $N_e$ would require
lower values of $\Lambda$, which push $v$ down, contrary to what we want.)
$\tau_3$ cannot be pushed above the Planck scale, so $a_i$ can only be decreased by at
most a factor of $10^{-3/4}$.  Although it also goes against the validity of the low-energy
effective supergravity theory upon which this model is based, in the interest of
being conservative when deriving constraints, 
we will consider values of the warp factor which are this small. 
By scanning over $g_s$ and $a_i$, we find the allowed and excluded regions
in the plane of the warped string scale and the string coupling, 
shown in figure (\ref{figkklmmt}).  One consequence is an upper bound on the
string coupling,
\beq
\label{gsc}
	g_s < 10^{-4.5}
\eeq
which is stringent, and smaller than normally expected.  
Our bound differs from
those found in  \cite{preheatNG2}, which also considered the nongaussian
perturbations produced by tachyonic preheating.  
Their allowed regions are also expressed in the plane of $M_s$ and $g_s$, but
differ markedly from ours.

A caveat to this result is the potential for complications due to the peculiar form
of the tachyon action (\ref{DBI})\footnote{The inflaton is also described by a DBI
action (though the potential does not appear as a prefactor), however, slow roll
ensures that it will always be a good approximation to expand the square root to
obtain a standard kinetic term.  In the case of the tachyon this may not always be a
good approximation, particularly during the  instability phase.}.  The dynamics of
the  open string tachyon action is different from the usual Klein-Gordon action
which our analysis assumes.  In the case of (\ref{DBI}) the potential is minimized
at $T = \infty$ and it takes an infinite amount of time for the condensation to
complete (and the quantum mechanical decay of the brane-antibrane system into closed
string states invalidates this classical description on a time scale comparable to
$(a_i M_s)^{-1}$).  Thus the criterion  $\langle(\delta\sigma)^2\rangle \sim v^2$
defines the limit of applicability of our analysis, rather than the end of the
symmetry breaking process.  Moreover, the field theory (\ref{DBI}) leads to
finite-time divergences in the spatial derivatives $\partial_i T$ near the core of
the topological defects where the field stays pinned at $T=0$ \cite{BBCS,div}, as
well as finite-time divergences in the second derivatives $\partial_i\partial_j T$
in regions where  $V \rightarrow 0$, called caustics \cite{caustics}.  It is not
clear how these peculiarities  might modify our estimate for $f_{NL}$, but we plan
to investigate this in future work.

\section{Conclusions}
\label{VIII}

We have carefully calculated the contribution from tachyonic preheating to the
spectrum and bispectrum of the curvature perturbation in a model of  hybrid
inflation, and derived constraints on the model parameters  using experimental
limits on the power spectrum and nongaussianity in the CMB.  The tachyonic
contribution to the spectrum, which we parametrized by $f_L$, has spectral index
$n=4$, and the experimental limit on $f_L$ turns out to always be a stronger
constraint than that coming from nongaussianity.  Thus we do not expect to see
nongaussian signals in future data sets, if hybrid inflation is the underlying
model.  Rather, one should look for an excess of power at small scales to see
the effects we have investigated.

We have studied a simplified model of hybrid inflation, in which the
tachyon is a real scalar field.  This model is not viable as it
stands, because domain walls will be copiously produced at the end of
inflation, and they will quickly dominate the universe.  We need to
assume that there are some small operators in the Lagrangian which violate
the symmetry $\sigma\to -\sigma$, and make the domain walls unstable.
Although our results are strictly valid only for hybrid inflation with real fields,
we have made a preliminary attempt to apply them to the warped brane-antibrane
inflation model, by matching the coefficients of the lowest dimension operators in 
the corresponding Lagrangians.  To the extent that this is valid, we find an
interesting limit on the string coupling, $g_s < 10^{-4.5}$.

One achievement of this work is our demonstration that nonlocal contributions to the
second order curvature perturbation, as defined herein, vanish when care is taken
to consistently include all terms of a given order in the slow-roll expansion.  The
nonlocal terms which appear at intermediate steps when using the generalized
longitudinal gauge have been the bane of some previous attempts to do similar
calculations.  They should not appear in any physical quantities since they violate
causality.  Any estimates made using the nonlocal terms are likely to grossly
overestimate perturbations on large scales, since these terms are singular at small
wave numbers.  

Some readers may feel uneasy about an event at the end of inflation
being able to affect fluctuations whose wavelength corresponds to
the horizon at the beginning of inflation.  However, as has been
emphasized in \cite{FB1,FB2}, there is no violation of causality 
here.  Consider a collection of comoving observers who are within the
horizon at the beginning of inflation, but are about to lose causal
contact with each other.  If they agree to set up perturbations with
small wavelengths in their Hubble patches at the end of inflation,
this will induce large-scale correlations between different Hubble
patches.  Since the tachyon field is synchronized across these different Hubble 
patches it is able to amplify the large-scale perturbations in a coherent way.

One avenue for future study is the more realistic class of models where $\sigma$ is
complex and the defects which form are cosmic strings instead of domain walls. 
However, this is a much more technically difficult model to study than the case of
the real scalar.  In the latter case, the VEV of $\sigma$ remains zero even during
the tachyonic preheating phase, because the field rolls in opposite directions on
either side of each domain wall, and the VEV averages to zero.  In the complex case,
the modulus of $\sigma$ grows uniformly during preheating, and the randomly varying
phase is only a gauge degree of freedom.  We can no longer decouple the equations
for the fluctuations of the metric and the inflationary fields when $\sigma$ has a
VEV.  It thus remains an interesting question whether the realistic hybrid inflation
model also gets significant nongaussianity from  tachyonic preheating
\cite{inprogress}.

\bigskip{\bf Acknowledgments.}  We would like to thank A.\ Mazumdar
and A. Jokinen for stimulating discussions which motivated us to
start on this work.  We also thank R.\ Brandenberger, C.\ Burgess,
D.\ Lyth, K.\ Malik,  A.\ Notari and D.\ Seery for helpful comments. 
We thank the referee for thoughtful and insightful comments which
helped us to strengthen our arguments.  This work was supported by
NSERC of  Canada and FQRNT of Qu\'ebec.

%%%%%%%%%%%%
%%%%%%%%%%%%

\renewcommand{\theequation}{A-\arabic{equation}}
\setcounter{equation}{0}
\section*{APPENDIX A: Generic Field in deSitter Space}
\label{deSitter}

In this appendix we review the solutions of the Klein-Gordon equation in deSitter space 
\cite{RiottoReview,VilenkinFord}.  If we consider the limit of deSitter space $\epsilon = \eta = 0$ then
$\varphi_0,H$ are constant and both the tachyon and the inflaton first order fluctuations satisfy an 
equation of the form
\begin{equation}
\label{xidS}
  \ddot{\xi}_k + 3 H \dot{\xi}_k + \left[ \frac{k^2}{a^2}  + m^2_{\xi} \right] \xi_k
   = 0
\end{equation}
or, in terms of conformal time
\[
  \xi''_k + 2 \sH \xi'_k + \left[ k^2 + a^2 m^2_{\xi} \right] \xi_k = 0
\]
where $a = e^{Ht} = -1/H\tau$ and $\sH = -1/\tau$.  Defining $f_k = a \xi_k$ the equation 
simplifies to
\begin{equation}
\label{massivedS}
  f''_k + \left[ k^2 + \frac{1}{\tau^2}\left(\frac{m_{\xi}^2}{H^2} - 2\right) \right] f_k = 0.
\end{equation}
This is the variable in terms of which the action is canonically normalized.  Choosing the Bunch-Davies 
vacuum corresponds to normalizing the solutions so that \cite{VilenkinFord}
\begin{equation}
\label{BDasympt}
  f_k(\tau) \approx e^{-ik\tau} / \sqrt{2 k}
\end{equation}
on small scales $k \gg a H$ or equivalently $-k\tau \gg 1$.
The general solutions to (\ref{massivedS}) are
\begin{equation}
\label{massiveSoln}
  f_k(\tau) = \sqrt{-\tau} \left[ c_1(k) H_{\nu}^\subo (-k\tau) +  c_2(k) H_{\nu}^\subt (-k\tau)\right]
\end{equation}
where the order of the Hankel functions is $\nu = \sqrt{9/4 - m^2_{\xi}/H^2}$.  This analysis makes no 
assumptions about the size of $m_{\xi}/H$ and all the formulae we derive are valid for arbitrary complex 
$\nu$ unless otherwise stated.  Notice that $-\tau \geq 0$ for all $t$ so that the arguments of the Hankel 
functions are always positive definite.  We recover the desired asymptotics (\ref{BDasympt})
on $-k\tau \ll 1$ with the choice
\[
  c_1(k) = \frac{\sqrt{\pi}}{2} \exp \left[\frac{i \pi}{2}\left( \nu + \frac{1}{2} \right) \right],
  \hspace{5mm} c_2(k) = 0.
\]
The solution becomes
\begin{equation}
\label{BDsoln}
  f_k(\tau)  = \frac{\sqrt{-\pi \tau}}{2} 
                     \exp \left[\frac{i \pi}{2}\left( \nu + \frac{1}{2} \right) \right]
                     H_{\nu}^\subo (-k \tau)
\end{equation}
or, in terms of cosmic time and the original field variable
\begin{equation}
\label{xiBDsoln}
  \xi_k(t) = \frac{1}{2} \sqrt{\frac{\pi}{a^3 H}} 
                      \exp \left[\frac{i \pi}{2}\left( \nu + \frac{1}{2} \right) \right]
                       H_{\nu}^\subo \left( \frac{k}{a H}\right).
\end{equation}
To get an intuitive feel for this initial condition we consider the asymptotic behaviour of (\ref{BDsoln}) 
on large scales:
\begin{eqnarray}
  f_k (\tau) &\rightarrow& \frac{\sqrt{-\pi \tau}}{2} 
                           \exp \left[\frac{i \pi}{2}\left( \nu + \frac{1}{2} \right) \right] \label{BDlargescale} \\
             &\times&         \left[
                               \frac{1}{\Gamma(\nu +1)} \left( -\frac{k \tau}{2}\right)^{\nu} -
                               i \frac{\Gamma(\nu)}{\pi} \left( -\frac{k \tau}{2}\right)^{-\nu}  
                               \right]\nonumber
\end{eqnarray}
as $- k\tau \rightarrow 0$.  To simplify further we should focus on either $m_{\xi}/H > 3/2$ or 
$m_{\xi}/H < 3/2$.

We consider a field which is heavy during inflation $m_{\xi} \gg H$ so that $\nu \approx i m_{\xi} / H$ is 
pure imaginary.  How do the two terms in the square braces of (\ref{BDlargescale}) compare?  The functions 
$(-k \tau)^{im_{\xi}/H}$ are oscillatory so the relative magnitude of these two terms depends only on the 
gamma function prefactors.  Using the results (for $\beta$ real and arbitrary complex $z$)
\[
  \Gamma(1 + z) = z \Gamma(z), \hspace{5mm}
  |\Gamma(1 + i \beta)| = \sqrt{\frac{\pi \beta}{\sinh (\pi \beta)}}
\]
from the theory of gamma functions we find that for $m_{\xi}/H \gg 1$
\beqa
  \left| \frac{1}{\Gamma(1 + i m_{\xi}/H)} \right| 
  \sim \sqrt{\frac{H}{2 \pi m_{\xi} }} \exp \left( \frac{\pi m_{\xi}}{2 H} \right),
  \hspace{2mm}\nonumber\\
  |\Gamma(i m_{\xi/}H)| \sim \sqrt{\frac{2 \pi H }{m_{\xi}}} \exp \left( -\frac{\pi m_{\xi}}{2 H} \right) \nonumber
\eeqa
so that the first term in the square braces on the second line of (\ref{BDlargescale}) dominates at large $m_{\xi}/H$.  Going 
back to the original variable we find 
\[
  |\xi_k| \sim a^{-3/2}\frac{1}{2^{3/2} \sqrt{ m_{\xi} }} 
  \hspace{5mm} \mathrm{for} \hspace{2mm} m_{\xi} \gg H
\]
on large scales.

We consider now a field which is light during inflation $m_{\xi} \ll H$.
In this case $\nu \approx 3/2$ is pure real and 
the first term in the square braces in (\ref{BDlargescale}) goes to zero.  We recover the scale invariant 
spectrum
\begin{equation}
\label{scaleinvariant}
  |\xi_k| \sim \frac{H}{\sqrt{2 k^3}} 
  \hspace{5mm} \mathrm{for} \hspace{2mm} m_{\xi} \ll H
\end{equation}
on large scales, as expected.

Notice that for any value of $m_{\xi}/H$ we have the small scale asymptotics
\[
  |\xi_k| \sim a^{-1} \frac{1}{\sqrt{2k}}
\]
by constuction.

During inflation $a(t)$ grows exponentially so that the large scale fluctuations of any massive field
are damped as $a^{-3/2}$ while the large scale fluctuations of any light field are approximately constant.  
On small scales the fluctuations of all fields get damped as $a^{-1}$.  

%%%%%%%%%%%%%%%%%%%%%%%%%%%%%%%%%%%%%%%%%%%%%%%%%%%%%%%%%%%%%%%%%%%%%%%%%%%%%%%%%%%%%%%%%%%

\renewcommand{\theequation}{B-\arabic{equation}}
\setcounter{equation}{0}
\section*{APPENDIX B:  Perturbed Einstein Equations and the Master Equation}
\label{masterapp}

Using Maple we have carefully verified the results of \cite{EV} for the perturbed Einstein equations and
the master equation.  Here we briefly review those results relevent for the computation of $\zeta^\subt$.
We present only the $\dtwo G^0_0 = \kappa^2 \dtwo T^0_0$, 
$\partial_i \dtwo G^i_0 = \kappa^2 \partial_i \dtwo T^i_0$ and 
$\delta^i_j \dtwo G^j_i = \kappa^2 \delta^i_j \dtwo T^j_i$ 
equations since the second order vector and
tensor fluctuations decouple from this system.  In the case that $\sigma_0 = 0$ the second order tachyon
fluctation $\dtwo \sigma$ decouples from the inflaton and gravitational fluctuations.  Analogously to the 
first order fluctuations, the Klein-Gordon equation for the inflaton fluctation at second order 
$\dtwo \varphi$ is not necessary to close the system of equations.  In the subsequent text we sometimes
insert the slow roll parameters $\epsilon$ and $\eta$ explicitly though we make no assumption that they are 
small.  We also introduce the shorthand notation $m_{\varphi}^2 = \partial^2 V / \partial \varphi^2$ and
$m_{\sigma}^2 = \partial^2 V / \partial \sigma^2$ and assume that
$\partial^2V/\partial\sigma\partial\varphi = 0$.

The second order $(0,0)$ equation
is
\begin{eqnarray}
&&  3 \sH \psi'^\subt  +(3-\epsilon)\sH^2 \phi^\subt - \Lap \psi^\subt \nonumber \\
&=& -\frac{\kappa^2}{2}  \left[ \varphi'_0 \dtwo \varphi' 
                      + a^2 \frac{\partial V}{\partial \varphi } \dtwo \varphi \right] + \Upsilon_1
\label{00}
\end{eqnarray}
where $\Upsilon_1$ is constructed entirely from first order quantities.  Dividing $\Upsilon_1$ into inflaton
and tachyon contributions we have
\[
  \Upsilon_1 = \Upsilon_1^{\varphi} + \Upsilon_1^{\sigma}
\]
where
\begin{eqnarray}
  \Upsilon_1^{\varphi} &=& 4(3-\epsilon)\sH^2\left(\phi^\subo\right)^2 
             + 2\kappa^2 \varphi'_0 \phi^\subo \done \varphi' \nonumber \\
             &-& \frac{\kappa^2}{2}\left( \done \varphi' \right)^2
             - \frac{\kappa^2}{2} a^2 m^2_{\varphi} \left( \done \varphi \right)^2 
             - \frac{\kappa^2}{2} \left(\Grad \done \varphi \right)^2 \nonumber \\
             &+& 8 \phi^\subo \Lap \phi^\subo
             + 3\left( \phi'^\subo \right)^2
             + 3\left( \Grad\phi^\subo\right)^2
\label{Upsilon1varphi}
\end{eqnarray}
and
\begin{eqnarray}
  \Upsilon_1^{\sigma} &=& - \frac{\kappa^2}{2} \left[ \left( \done \sigma'\right)^2 
  + \left( \Grad \done \sigma\right)^2 \right. \nonumber \\ 
  &+& \left. a^2 m_{\sigma}^2 \left(\done \sigma \right)^2 \right]. 
\label{Upsilon1sigma}
\end{eqnarray}
The divergence of the second order $(0,i)$ equation is
\begin{equation}
\label{0i}
  \Lap \left[ \psi'^\subt + \sH \phi^\subt \right] = 
  \frac{\kappa^2}{2} \varphi'_0 \Lap \dtwo \varphi + \Upsilon_2
\end{equation}
where $\Upsilon_2 = \Upsilon_2^{\varphi} + \Upsilon_2^{\sigma}$ is constructed entirely from first order 
quantities.  The inflaton part is
\begin{eqnarray}
  \Upsilon_2^{\varphi} 
             &=& 2\kappa^2 \varphi'_0 \partial_i \left( \phi^\subo \partial^i \done \varphi \right)
             + \kappa^2 \partial_i \left( \done \varphi' \partial^i \done \varphi \right) \nonumber \\
             &-& 8 \partial_i \left( \phi^\subo \partial^i \phi'^\subo \right)
             - 2 \partial_i \left( \phi'^\subo \partial^i \phi^\subo \right)
\label{Upsilon2varphi}
\end{eqnarray}
and the tachyon part is
\begin{equation}
\label{Upsilon2sigma}
  \Upsilon_2^{\sigma} = \kappa^2 \partial_i \left( \done \sigma' \partial^i \done \sigma \right).
\end{equation}
The trace of the second order $(i,j)$ equation is
\begin{eqnarray}
  && 3\psi''^\subt + \Lap \left[ \phi^\subt - \psi^\subt \right] + 6 \sH \psi'^\subt \nonumber \\
  &+& 3 \sH \phi'^\subt + 3(3-\epsilon) \sH^2 \phi^\subt \nonumber \\
  &=& \frac{3\kappa^2}{2}\left[  \varphi'_0 \dtwo \varphi' 
      - a^2 \frac{\partial V}{\partial \varphi } \dtwo \varphi\right] + \Upsilon_3
\label{ij}
\end{eqnarray}
where $\Upsilon_3 = \Upsilon_3^{\varphi} + \Upsilon_3^{\sigma}$ is constructed entirely from first 
order quantities.  The inflaton part is
\begin{eqnarray}
  \Upsilon_3^{\varphi} &=& 12(3-\epsilon)\sH^2 \left( \phi^\subo\right)^2 
             - 6 \kappa^2 \varphi'_0 \phi^\subo \done\varphi' \nonumber \\
             &+& \frac{3\kappa^2}{2}\left(\done \varphi'\right)^2 
             - \frac{3 \kappa^2}{2} a^2 m^2_{\varphi} \left( \done \varphi \right)^2
             - \frac{\kappa^2}{2}\left(\Grad \done \varphi \right)^2 \nonumber \\
             &+& 3 \left( \phi'^\subo\right)^2 
             + 8 \phi^\subo \Lap \phi^\subo
             + 24 \sH \phi^\subo \phi'^\subo \nonumber \\
             &+& 7 \left( \Grad \phi^\subo \right)^2
\label{Upsilon3varphi}
\end{eqnarray}
and the tachyon part is
\begin{eqnarray}
  \Upsilon_3^{\sigma} &=& \kappa^2 \left[ \frac{3}{2}\left( \done \sigma' \right)^2
                      - \frac{1}{2} \left(\Grad\done\sigma\right)^2 \right. \nonumber \\
                      &-& \left. \frac{3}{2}a^2 m_{\sigma}^2 \left( \done \sigma \right)^2 \right]
\label{Upsilon3sigma}
\end{eqnarray}

We now proceed to derive the master equation.  Adding (\ref{00}) to the inverse laplacian of the time 
derivative of (\ref{0i}) and then using (\ref{0i}) to eliminate $\dtwo \varphi$ yields
\begin{eqnarray}
  &&\psi''^{\subt} - (1 + 2\epsilon - 2\eta)\sH \psi'^\subt + \sH \phi'^\subt \nonumber \\
  &-& 2(2\epsilon-\eta)\sH^2 \phi^\subt - \Lap \psi^\subt = \Upsilon_1 + \Linv \Upsilon'_2 \nonumber \\
  &-& 2(2+\epsilon-\eta)\sH\Linv\Upsilon_2.
\label{noinflaton}
\end{eqnarray}
Notice that we have decoupled $\psi^\subt$ and $\phi^\subt$ from the inflaton perturbation $\dtwo \varphi$.
It remains now to express $\psi^{\subt}$ in terms of $\phi^\subt$.  To this end we subtract the inverse 
laplacian of (\ref{0i}) from (\ref{ij}) and again use (\ref{0i}) to eliminate $\dtwo \varphi$ which gives
\[
  \Lap \left[ \phi^\subt - \psi^\subt \right] = \Upsilon_3 - 3 \Linv \Upsilon'_2 - 6 \sH \Linv \Upsilon_2
\]
or, equivalently
\begin{equation}
\label{phipsi}
  \psi^\subt = \phi^\subt - \Linv \gamma.
\end{equation}
Following the notation of \cite{EV} we have defined
\begin{equation}
\label{gam}
  \gamma = \Upsilon_3 - 3 \Linv \Upsilon'_2 - 6 \sH \Linv \Upsilon_2
\end{equation}
which can be split into inflaton and tachyon components $\gamma = \gamma_{\varphi} + \gamma_{\sigma}$ in
an obvious fashion.  Our $\Upsilon_2$ is related to the quantities $\alpha,\beta$ defined in \cite{EV} 
by
\[
  \Upsilon_2 = \kappa^2 \beta - \alpha.
\]
Now, using (\ref{phipsi}) to eliminate $\psi^\subt$ from (\ref{noinflaton}) gives the master equation
\begin{eqnarray*}
  &&\phi''^\subt + 2(\eta-\epsilon)\sH \phi'^\subt 
  + \left[ 2(\eta-2\epsilon)\sH^2 - \Lap \right] \phi^\subt \\
  &=& \Upsilon_1 + \Linv \Upsilon'_2 - 2(2+\epsilon-\eta)\sH\Linv\Upsilon_2 - \gamma \\
  &-& (1 + 2\epsilon - 2\eta)\sH \Linv\gamma' + \Linv\gamma''
\end{eqnarray*}
Inserting explicitly the expression for $\gamma$ (\ref{gam}) this can be written as
\[
  \phi''^\subt + 2(\eta-\epsilon)\sH \phi'^\subt 
  + \left[ 2(\eta-2\epsilon)\sH^2 - \Lap \right] \phi^\subt = J
\]
where the source is
\begin{eqnarray}
  J &=& \Upsilon_1 - \Upsilon_3 + 4\Linv\Upsilon'_2 + 2(1-\epsilon+\eta)\sH \Linv\Upsilon_2 \nonumber \\ 
    &+& \Linv \gamma'' - (1 + 2\epsilon-2\eta)\sH\Linv\gamma'. \label{appsource}
\end{eqnarray}
We can split the source into tachyon and inflaton contributions $J = J^{\varphi} + J^{\sigma}$ in the obvious
manner, by taking the tachyon and inflaton parts of $\Upsilon_1,\Upsilon_2,\Upsilon_3,\gamma$.

We now derive some results concering the tachyon source terms which will be useful in the
text.  First we consider the tachyon contribution to $\gamma$ (\ref{gam}):
\[
  \gamma_\sigma = \Upsilon_3^\sigma - 3\Linv\partial_\tau\Upsilon_2^\sigma
  - 6\sH\Linv\Upsilon_2^\sigma.
\]
Using equations (\ref{Upsilon2sigma}) and (\ref{Upsilon3sigma}) we can write this as
\begin{eqnarray}
   \gamma_\sigma &=& \kappa^2\left[\frac{3}{2}(\done\sigma')^2 
   - \frac{1}{2}\partial_i\done\sigma\partial^i\done\sigma
   - \frac{3}{2}a^2m_\sigma^2(\done\sigma)^2 \right] \nonumber \\
   &-&3\kappa^2 \Linv\partial_\tau\partial_i(\done\sigma'\partial^i\done\sigma) \nonumber \\
   &-& 6\kappa^2 \sH\Linv\partial_i(\done\sigma'\partial^i\done\sigma) \label{gamma_sigma1}
\end{eqnarray}
We now write $\gamma_\sigma$ as
\begin{eqnarray*}
   \gamma_\sigma &=& \kappa^2\Linv\left[\frac{3}{2}\Lap(\done\sigma')^2 
   - \frac{1}{2}\Lap(\partial_i\done\sigma\partial^i\done\sigma) \right. \\
   &-& \frac{3}{2}a^2m_\sigma^2\Lap(\done\sigma)^2 
   -3\partial_\tau\partial_i(\done\sigma'\partial^i\done\sigma) \\
   &-& \left. 6\sH\partial_i(\done\sigma'\partial^i\done\sigma) \right]
\end{eqnarray*}
and, after some algebra, we have
\begin{eqnarray*}
   \gamma_\sigma &=& \kappa^2\Linv\left[ 
   -\frac{1}{2}\Lap(\partial_i\done\sigma\partial^i\done\sigma) \right. \\
   &-& 3 \partial_i\left(\done\sigma'' +  2\sH\done\sigma' + a^2m_\sigma^2\done\sigma\right)
   \partial^i\done\sigma \\
   &-& \left. 3 \left(\done\sigma''  + 2\sH\done\sigma' + a^2m_\sigma^2\done\sigma\right)
   \Lap\done\sigma  \right]
\end{eqnarray*}
The last two lines can be simplified using the equation of motion for the tachyon 
fluctuation
\[
  \done \sigma'' + 2\sH \done \sigma + a^2 m_\sigma^2 \done \sigma = \Lap\done\sigma
\]
which gives
\begin{eqnarray}
   \gamma_\sigma &=& \kappa^2\Linv\left[ 
   -\frac{1}{2}\Lap\left(\partial_i\done\sigma\partial^i\done\sigma\right) \right. \nonumber
    \\
   &-&\left.  3\partial_i(\Lap\done\sigma)\partial^i\done\sigma 
   - 3 \partial_i\partial^i\done\sigma \Lap \done\sigma \right] \nonumber \\
  &=&  \kappa^2\Linv\left[ 
   -\frac{1}{2}\Lap\left(\partial_i\done\sigma\partial^i\done\sigma\right) \right. 
  \nonumber \\
   &-&\left.  3\partial_i\left(\Lap\done\sigma\partial^i\done\sigma\right) \right]
\label{gamma_sigma2}
\end{eqnarray}
This result has also been derived in \cite{EV}.  We now consider the tachyon contribution
to the source.  We take
\begin{eqnarray*}
  J^\sigma &=& \Upsilon_1^\sigma - \Upsilon_3^\sigma + 4\Linv\partial_\tau\Upsilon_2^\sigma
   +
  2(1-\epsilon + \eta)\sH\Linv\Upsilon_2^\sigma \\
  &+& \Linv\gamma_\sigma'' - (1 + 2\epsilon - 2\eta)\sH\Linv\gamma_\sigma'
\end{eqnarray*}
and, using (\ref{Upsilon1sigma}), (\ref{Upsilon2sigma}) and (\ref{Upsilon3sigma}), we have
\begin{eqnarray}
  J^\sigma(\tau,\vec{x}) &=& a^2 \kappa^2 m_{\sigma}^2 \left( \done \sigma \right)^2
                               - 2 \kappa^2 \left( \done \sigma' \right)^2 \nonumber \\
  &+& 2 \kappa^2 \sH (1+\eta-\epsilon) \Linv \partial_i \left( \done \sigma' \partial^i \done \sigma \right)
  \nonumber\\ &+& 4 \kappa^2 \Linv \partial_{\tau} \partial_i \left( \done \sigma' \partial^i \done \sigma \right) 
  \nonumber \\
  &-& \sH (1 + 2\epsilon - 2\eta) \Linv \gamma'_{\sigma} + \Linv \gamma''_{\sigma}.
  \label{Tsource}
\end{eqnarray}

%%%%%%%%%%%%%%%%%%%%%%%%%%%%%%%%%%%%%%%%%%%%%%%%%%%%%%%%%%%%%%%%%%%%%%%%%%%%%%%%%%%%%%%%%%%%

\renewcommand{\theequation}{C-\arabic{equation}}
\setcounter{equation}{0}
\section*{APPENDIX C:  The Inflaton Contribution to $\zeta^\subt$}

In this appendix we consider the calculation of the inflaton part of the second order
curvature perturbation $\zeta^\subt_{\varphi}$ using results from appendix B and following closely the 
calculation of the tachyon part of the second order curvature perturbation which is discussed in subsection
\ref{tachyonPart}.  

The first step to computing $\zeta^\subt_{\varphi}$ is to use the Green function (\ref{green}) to eliminate
the dependence of $\zeta^\subt_{\varphi}$ on $\phi^\subt$ (from the first line of (\ref{zeta2})
which is equivalent to (\ref{zeta2firstline})) in favour of $\phi^\subo$, $\done\varphi$.  This procedure 
follows closely the analysis of subsection \ref{tachyonPart} and in particular (\ref{subleadinggreen}) 
still holds though here we are interested only in the inflaton part of the soucre $J^{\varphi}$.

Having removed the explicit dependence of $\zeta^\subt$ on the second
order metric perturbation $\phi^\subt$ we next eliminate
$\phi^\subo,\done\varphi$ in favour of $\zeta^\subo$. Using the
solution (\ref{infPhi1}) and the first order constraint equation
(\ref{(0,i)}) one may verify that on large scales
\[
  \zeta^\subo \cong -\phi^\subo - \frac{\varphi'_0}{\sH} \done \varphi 
               \cong -\frac{1}{\epsilon} \phi^\subo
\]
and
\[
  \done \varphi \cong \frac{\varphi'_0}{\sH}\frac{1}{\epsilon} \phi^\subo
                \cong -\frac{\varphi'_0}{\sH}\zeta^\subo.
\]
On large scales the first order curvature perturbation is approximately constant since \cite{BellidoWands}
\[
  \zeta'^\subo = \frac{\Lap}{\epsilon \sH} \phi^\subo
\]
using the fact that $\sigma_0=0$ (so that there are no anisotropic
stresses, whose absence guarantees the conservation of 
$\zeta^\subo$ on super-Hubble scales).
Thus on large scales we have
\[
  \done \varphi' \cong -(2\epsilon-\eta)\varphi'_0\zeta^\subo.
\]

It is straightforward to compute the last three lines of
(\ref{zeta2}) using the fact that $Q^\subo_{\varphi} \approx -\varphi'_0 \zeta^\subo / \sH$ and 
$Q'^\subo_{\varphi} \approx -(2\epsilon - \eta) \varphi'_0 \zeta^\subo$ on large scales.  The result is
\begin{equation}
\label{easypart}
  \zeta^\subt \ni \left[ \frac{1}{3}\left(\frac{a \, m_{\varphi}}{\sH}\right)^2 
              + 2 + 2\epsilon - 2\eta \right] (\zeta^\subo)^2
\end{equation}
where we have dropped terms which are higher order in slow roll parameters or which contain gradients.  
Notice that the quantity $(a \, m_{\varphi}/ \sH)^2$ is first order in the slow roll expansion because
\[
  \eta \approx \frac{1}{\kappa^2}\frac{1}{V}\frac{\partial^2 V}{\partial \varphi^2} 
       = \frac{1}{\kappa^2}\frac{m_{\varphi}^2}{V}
       \approx \frac{1}{3}\left(\frac{a \,  m_{\varphi}}{\sH}\right)^2.
\]
To (\ref{easypart}) we must add the contribution coming from the fist line of (\ref{zeta2}) which can be
written explicitly in terms of first order quantities using equations (\ref{zeta2firstline}) 
and (\ref{subleadinggreen}).  Here we consider only the particular solution for $\phi^\subt$ due to the 
inflaton source $J^{\varphi}$.

In order to compute $J^{\varphi}$ we first study the quantities 
$\Upsilon_1,\Upsilon_2,\Upsilon_3,\Upsilon_4,\gamma$ which are defined in appendix B.
On large scales and to leading order in slow roll we have
\[
  \Upsilon_1^{\varphi} \approx \left[ 12\epsilon^2 
  - \epsilon\left(\frac{a \, m_{\varphi}}{\sH}\right)^2 \right] \sH^2 (\zeta^\subo)^2.
\]
The quantity $\Linv \Upsilon_2^{\varphi}$ can be written as (see equation 39 of \cite{EV})
\begin{eqnarray*}
  \Linv \Upsilon_2^{\varphi} &=& \frac{\kappa^2}{2}(\epsilon-\eta) \sH (\done \varphi)^2 
                                 + 3\sH (\phi^\subo)^2 - 2 \phi^\subo \phi'^\subo \\
                             &+& \frac{2}{\varphi'_0}
                                 \Linv\left(\partial_i\Lap\phi^\subo\partial^i\done\varphi 
                                  \right. \nonumber \\
                             &+& \left. \Lap \phi^\subo \partial_i\partial^i \done \varphi \right).
\end{eqnarray*}
The inverse laplacians on the last two lines contribute only to the momentum dependence of $f_{NL}^{\varphi}$
which we neglect.  On large scales and in the slow roll limit we have
\[
  \Linv \Upsilon_2^{\varphi} \approx (4\epsilon^2 - \epsilon\eta)\sH (\zeta^\subo)^2 + \cdots
\]
where the $\cdots$ denotes momentum dependent terms.  On large scales and in the slow roll limit we also
have
\[
  \Upsilon_3^{\varphi} \approx \left[ 36 \epsilon^2 
  - 3\epsilon\left(\frac{a\,m_{\varphi}}{\sH}\right)^2 \right] \sH^2 (\zeta^\subo)^2.
\]
The quantity $\Upsilon_4$ defined in (\ref{Upsilon4}) depends only on the inflaton fluctuation and is given 
by 
\[
  \Upsilon_4 = \Upsilon_4^{\varphi} \approx 
  (6\epsilon^2 - 2\epsilon\eta)\frac{\varphi'_0}{\sH}(\zeta^\subo)^2
\]
on large scales and in the slow roll limit.
Using these results one may readily verify that $\gamma_{\varphi}/\sH^2$ is third order in slow roll 
parameters
\[
  \gamma_{\varphi} \approx \mathcal{O}(\epsilon^3) \sH^2 (\zeta^\subo)^2.
\]

Using these results one may verify that the only term in (\ref{zeta2firstline}) which contributes at lowest 
order in slow roll parameters is the term proportional to $\Upsilon_1^{\varphi}/\epsilon\sH^2$.   
Thus, the contribution to the second order
curvature perturbation due to the first line of (\ref{zeta2}) is
\[
  \zeta^\subt_{\varphi} \ni \left[ 4\epsilon 
  - \frac{1}{3}\left(\frac{a\, m_{\varphi}}{\sH}\right)^2\right](\zeta^\subo)^2.
\]

Adding all the contributions together we find
\[
  \zeta^\subt_{\varphi} \approx (2 - 2\eta + 6\epsilon)(\zeta^\subo)^2.
\]
The contribution $2(\zeta^\subo)^2$ stems from using the Malik and Wands \cite{MW}
definition of the second order curvature perturbation.  
It can be related to the definition of 
Lyth and Rodriguez \cite{lythNG} 
(which also agrees with Maldacena \cite{Maldacena}) using 
\[
  \zeta^\subt = \zeta^\subt_{LR} + 2(\zeta^\subo)^2.
\]
The Lyth-Rodriguez curvature perturbation, due to the inflaton up to 
second order, can thus be written as
\[
  \zeta^{\varphi}_{LR} = \zeta^\subo - \frac{3}{5}f_{NL}^{\varphi} (\zeta^\subo)^2
\]
where
\[
  f_{NL}^{\varphi} = \frac{5}{6}(2\eta-6\epsilon).
\]
In writing $\zeta^{\varphi}_{LR}$ above we have suppressed the 
homogeneous $k=0$ mode of $\zeta$ which should
be subtracted to ensure that $\langle \zeta \rangle = 0$.
This result differs from previous studies \cite{Maldacena,SeeryLidsey} by a factor
of two.  The calculation of \cite{Maldacena,SeeryLidsey} takes into account the effect of nonlinear
evolution aswell as the effect of computing the bispectrum in the vacuum of the interacting theory,
as opposed to the vacuum of the free theory.  Our calculation does not consider the effect of the change
in vacuum which is the same order of magnitude.  Thus we should not expect to reproduce exactly the
results of \cite{Maldacena,SeeryLidsey}.  The change in vacuum will not change the calculation of the 
tachyon part of the curvature perturbation $\zeta^\subt_{\sigma}$ since $\zeta^\subo$ does not depend
on $\done \sigma$ and hence contributions to $\zeta_{\sigma}$ due to the change in vacuum will be higher
than second order in perturbation theory.

%%%%%%%%%%%%%%%%%%%%%%%%%%%%

\renewcommand{\theequation}{D-\arabic{equation}}
\setcounter{equation}{0}
\section*{APPENDIX D: Fourier Transforms, Mode Functions and Inverse Laplacians}
\label{fouriertransforms}

We define the Fourier transform of some first order quantity $\delta f(t,\vec{x})$ 
by
\begin{eqnarray}
  \delta f(t,\vec{x}) &=& \int \frac{d^3 k}{(2 \pi)^{3/2}} e^{i \vec{k} \cdot \vec{x}} 
                         \delta\tilde{f}_k(t) \label{xform} \\
  \delta \tilde{f}_{\vec{k}}(t) &=& \int \frac{d^3 x}{(2 \pi)^{3/2}} e^{- i \vec{k} \cdot \vec{x} } 
                             \delta f(t,\vec{x}) \nonumber
\end{eqnarray}
where the Hermiticity of $\delta f(t,\vec{x})$ implies that $\delta \tilde{f}_k(t) = \delta\tilde{f}_{-k}(t)^{\dagger}$ 
so that we can define
\begin{equation}
\label{modefn}
  \delta\tilde{f}_k(t) = a_{k} \xi_{k}(t) + a_{-k}^\dagger \xi_{-k}(t)^{\star}
\end{equation}
where $a_k$ is an operator and $\delta f_k(t)$ is a c-number valued mode function.  We then re-write the 
Fourier transform as
\begin{eqnarray*}
  \delta f(t,x) &=& \int \frac{d^3 k}{(2\pi)^{3/2}} \left[ a_k \xi_k(t)e^{ikx} 
                                               + a_{-k}^\dagger \xi_{-k}(t)^{\star}e^{ikx} \right] \\
                &=& \int  \frac{d^3 k}{(2\pi)^{3/2}} \left[ a_k \xi_k(t) e^{ikx} 
                                                   + a_k^{\dagger} \xi_k(t)^{\star} e^{-ikx} \right].
\end{eqnarray*}
In this form it is clear that $a_k$ is the usual creation operator satisfying
\[
  \left[ a_{k}, a_{k'}^{\dagger} \right] = \delta^{(3)}(\vec{k} - \vec{k'})
\]
and one should expect
\[
  \xi_k(\tau) \approx \frac{1}{\sqrt{2 k}} e^{- i k \tau}
\]
on small scales, which corresponds to the Bunch-Davies vacuum $a_k |0\rangle = 0$.  This is 
consistent with the usual definition of the power spectrum in terms of the two-point function
\begin{eqnarray*}
\left\langle0|\delta f^2(t,\vec{x})|0\right\rangle 
                       &=& \int \frac{d^{\,3}k}{(2\pi)^{3}} |\xi_k(t)|^2 \\
                       &=& \int \frac{dk}{k} \mathcal{P}_f(k)
\end{eqnarray*}
so that
\[
  \mathcal{P}_f(k) = \frac{k^3}{2 \pi^2} |\xi_k(t)|^2.
\]

We now discuss the Fourier transform of the tachyon source.  Typical terms in the source have the form
\begin{eqnarray*}
  J^\subo(t,\vec{x}) &=& b(t) \delta f(t,\vec{x}) \delta g(t,\vec{x}), \\
  J^\subt(t,\vec{x}) &=& b(t) \triangle^{-1} \left[ \delta f(t,\vec{x}) \delta g(t,\vec{x}) \right], \\
 J^\subth(t,\vec{x}) &=& b(t) \triangle^{-2} \left[ \delta f(t,\vec{x}) \delta g(t,\vec{x}) \right]
\end{eqnarray*}
where $\delta f$, $\delta g$ are some first order quantities and $b$ is constructed from zeroth order
quantities.  
The Green's function for the laplacian is defined as appropriate in the absence of boundary surfaces
\[
  (\triangle^{-1} f)(t,\vec{x}) = \int d^3 x' G(\vec{x} - \vec{x}') f(t,\vec{x}') 
\]
where
\[
  G(\vec{x} - \vec{x}') = -\frac{1}{4 \pi} \frac{1}{|\vec{x} - \vec{x}'|}.
\]
In Fourier space we have
\begin{eqnarray*}
  G(x,x') &=& \int \frac{d^{\,3}k}{(2\pi)^{3/2}} e^{ikx} G_k(x') \\
  G_k(x') &=&  - \frac{1}{(2 \pi)^{3/2}} \frac{e^{-ikx'}}{k^2}.
\end{eqnarray*}
In Fourier space the souce terms contain convolutions
\begin{eqnarray*}
  J^\subo_k(t) &=& b(t) (\delta\tilde{f} \star \delta\tilde{g})_k(t), \\
  J^\subt_k(t) &=& - b(t) \frac{1}{k^2} (\delta\tilde{f} \star \delta\tilde{g})_k(t), \\
 J^\subth_k(t) &=& b(t) \frac{1}{k^4}(\delta\tilde{f} \star \delta\tilde{g})_k(t).
\end{eqnarray*}
where $\tilde{f}_k$, $\tilde{g}_k$ are operator valued Fourier transforms defined as in (\ref{xform}).  
These may be related to the mode functions as (\ref{modefn}).  Finally, we have defined 
convolution by
\[
  (\delta\tilde{f} \star \delta\tilde{g})_{\vec{k}}(t) = 
  \int \frac{d^3 k'}{(2\pi)^{3/2}} \delta\tilde{f}_{\vec{k}'}(t) \delta\tilde{g}_{\vec{k} - \vec{k}'}(t) 
\]

%%%%%%%%%%%%%%%%%%%%%%%%%%%%%%%%%%%%%%%%%%%%%%%%%%%%%%%%%%%%%%%%%%%%%%%%%%%%%%%%%%%%%%%%

\renewcommand{\theequation}{E-\arabic{equation}}
\setcounter{equation}{0}
\section*{APPENDIX E: Construction of The Tachyon Curvature Perturbation}
%\label{tcurv}

In this appendix we include technical details about the contruction of
$\zeta^\subt_\sigma$ using the Green function for the master equation.
For clarity we repeat some details which are included in the text.  We consider
only contributions to the equations which depend on the tachyon field and we
remind the reader that $\zeta^\subo$ is independent of $\sigma$.  The
second order curvature perturbation is (see (\ref{zeta2}-\ref{Q1sigma}))
\begin{eqnarray}
 \zeta^\subt &=& 
  \frac{1}{3-\epsilon} \frac{1}{(\varphi'_0)^2}\left[ \varphi'_0 Q'^\subt_{\varphi} 
   + a^2 \frac{\partial V}{\partial \varphi} Q^\subt_{\varphi} \right] \nonumber \\
  &+& \frac{1}{3-\epsilon} \frac{1}{(\varphi'_0)^2}\left[ \left(\done\sigma'\right)^2 
                 + a^2 m_{\sigma}^2 \left(\done\sigma\right)^2 \right] \nonumber \\
             &+& \mathrm{inflaton \hspace{2mm} contributions}
\label{zeta2_app}
\end{eqnarray}
where the second order Sasaki-Mukhanov variable is
\begin{eqnarray}
  Q^\subt_{\varphi} &=& \dtwo \varphi + \frac{\varphi'_0}{\sH}\psi^\subt \nonumber \\
                    &+& \mathrm{inflaton \hspace{2mm} contributions}
\label{Q2varphi_app}
\end{eqnarray}
Inserting (\ref{Q2varphi_app}) into (\ref{zeta2_app}) and using the $(0,0)$ Einstein
equation (\ref{00}) to eliminate the contribution 
$\varphi_0'\dtwo \varphi' + a^2 \partial V/\partial\varphi\dtwo\, \varphi$ gives
\begin{eqnarray}
  \zeta^\subt_\sigma 
  &=& -\frac{\phi'^\subt}{\epsilon \sH}  - \left( \frac{1}{\epsilon} + 1 \right) \phi^\subt
  + \frac{1}{3-\epsilon}\frac{\Lap \phi^\subt}{\epsilon \sH^2} \nonumber \\
  &+& \frac{1}{\epsilon \sH} \Linv \gamma_\sigma' + \Linv \gamma_\sigma 
   - \frac{1}{3-\epsilon} \frac{1}{\epsilon \sH^2} \gamma_\sigma
\label{zeta2_tachyon}
\end{eqnarray}
where we have also eliminated $\psi^\subt$ in favour of $\phi^\subt$ using (\ref{phipsi}).
Notice that using (\ref{00}) introduces a term proportional to $\Upsilon_1^\sigma$ to
the curvature perturbation which cancels the contribution proportional to 
$(\done\sigma')^2 + a^2m_\sigma^2(\done\sigma)^2$ on the second line of (\ref{zeta2_app})
up to a gradient term $(\Grad\done\sigma)^2$ which can be neglected
on large scales.  The result (\ref{zeta2_tachyon}) is valid only on large scales but we
have not yet assumed slow roll.
In (\ref{zeta2_tachyon}) it is understood that $\phi^\subt$ denotes only the particular 
solution due to the tachyon source, $J^\sigma$.  We now use the Green function 
(\ref{green}) to solve for $\phi^\subt$ in terms of the tachyon source (\ref{Tsource}).  
We work only to leading order in slow roll parameters.  We also work in the large scale 
limit.  To lowest order in $\epsilon$, $\eta$ and up to order $k^2$ we can write
\begin{eqnarray}
  &-& \frac{\phi'^\subt_k}{\epsilon\sH} - \frac{\phi^\subt_k}{\epsilon} 
  - \frac{k^2\phi^\subt_k}{3\epsilon\sH^2} = 
  \nonumber \\
  &&\frac{1}{\epsilon}\int_{\tau_i}^0d\tau' \Theta(\tau-\tau')J_k^{\sigma}(\tau') \nonumber \\
  &\times& \left[
  \tau' + \left(-\frac{1}{6}\tau'^3 + \frac{5}{6}\tau^2\tau' - \frac{2}{3}\tau^3\right) k^2   \right] 
  \label{subleadinggreen_app}
\end{eqnarray}
using the Green function (\ref{green}) and the relation (\ref{particular}).  
Equation 
(\ref{subleadinggreen_app}) allows us to eliminate the dependence of $\zeta^\subt_\sigma$
on $\phi^\subt$.  At leading
order in slow roll parameters the tachyon source is (see (\ref{appsource}))
\begin{eqnarray}
  J^\sigma &=& \Upsilon_1^\sigma - \Upsilon_3^\sigma + 4\Linv\partial_\tau\Upsilon^\sigma_2 
  + 2\sH \Linv\Upsilon_2^\sigma \nonumber \\ 
  &+& \Linv\partial_\tau^2 \gamma_\sigma - \sH\Linv\partial_\tau\gamma_\sigma. 
  \label{leadingsource_app}
\end{eqnarray}

We must now insert (\ref{leadingsource_app}) into (\ref{subleadinggreen_app}), perform numerous
integrations by parts, and then insert this result into (\ref{zeta2_tachyon}).  We evaluate
term-by-term the last line of (\ref{subleadinggreen_app}).  The first contribution 
(the term proportional to $k^0$) is
\[
  - \frac{\phi'^\subt_k}{\epsilon\sH} - \frac{\phi^\subt_k}{\epsilon} 
  - \frac{k^2\phi^\subt_k}{3\epsilon\sH^2} \ni 
  -\frac{1}{\epsilon}\int_{\tau_i}^{\tau}d\tau' \frac{J^\sigma(\tau')}{\sH(\tau')}.
\]
Inserting (\ref{leadingsource_app}), noting that $\sH(\tau') = -1/\tau'$ at leading order
in slow
roll and integrating by parts gives
\begin{eqnarray}
 &-& \frac{\phi'^\subt_k}{\epsilon\sH} - \frac{\phi^\subt_k}{\epsilon} 
  - \frac{k^2\phi^\subt_k}{3\epsilon\sH^2} \ni 
 -\frac{1}{\epsilon}\int_{\tau_i}^{\tau}d\tau' \frac{J^\sigma(\tau')}{\sH(\tau')} 
 \nonumber \\
 &\cong& \frac{1}{\epsilon}\int_{\tau_i}^{\tau}d\tau'
 \left[ -\frac{\Upsilon_1^\sigma}{\sH(\tau')} 
 + \frac{\Upsilon_3^\sigma}{\sH(\tau')} - 6\Linv\Upsilon_2^\sigma  \right] 
 \nonumber \\
 &+& \frac{1}{\epsilon}\left[-4\frac{\Linv\Upsilon_2^\sigma}{\sH} - 
 \frac{\Linv\gamma'_\sigma}{\sH} \right] + \cdots \label{first_app}
\end{eqnarray}
where the terms under the $d\tau'$ integral are evaluated at $\tau'$ while the terms in the 
square braces on the second line are evaluated at $\tau$.  The $\cdots$ denotes constant 
terms
evaluated at $\tau = \tau_i$ which arise from the integration by parts.  Since our interest
is in the preheating phase during which the fluctuations $\done\sigma$ are amplified 
exponentially we can safely drop these constant terms.  

The second contribution to the last line of (\ref{subleadinggreen_app}) has the form
\[
 - \frac{\phi'^\subt_k}{\epsilon\sH} - \frac{\phi^\subt_k}{\epsilon} 
  - \frac{k^2\phi^\subt_k}{3\epsilon\sH^2} \ni
 \frac{1}{6\epsilon}\int_{\tau_i}^{\tau} d\tau'\frac{k^2J_k^\sigma}{\sH(\tau')^3}.
\]
In evaluating this we need only consider terms in $k^2 J_k^\sigma$ which are not suppressed
on large scales.  Using (\ref{leadingsource_app}) one may verify that
\begin{equation}
\label{large_k^2J_app}
  k^2 J_k^\sigma \cong -\gamma''_{\sigma,k} + \sH \gamma'_{\sigma,k}
\end{equation}
on large scales (recall that $\Upsilon_2^\sigma$ is a gradient, see (\ref{Upsilon2sigma}))
so that we have
\begin{eqnarray}
 &-& \frac{\phi'^\subt_k}{\epsilon\sH} - \frac{\phi^\subt_k}{\epsilon} 
  - \frac{k^2\phi^\subt_k}{3\epsilon\sH^2} \ni
\frac{1}{6\epsilon}\int_{\tau_i}^{\tau} d\tau'\frac{k^2J_k^\sigma}{\sH(\tau')^3} 
 \nonumber \\
 &\cong& \frac{1}{\epsilon}\int_{\tau_i}^{\tau}d\tau'\left[
 -\frac{2}{3}\frac{\Upsilon_3^\sigma}{\sH} + 6\Linv\Upsilon_2^\sigma \right] \nonumber \\
 &+& \left[-\frac{1}{6\epsilon}\frac{\gamma'_\sigma}{\sH^3} - 
 \frac{1}{3\epsilon}\frac{\gamma_\sigma}{\sH^2}
 + 2\frac{\Linv\Upsilon_2^\sigma}{\sH}\right] + \cdots \label{second_app}
\end{eqnarray}
Equation (\ref{large_k^2J_app}) shows why it was necessary to include the $k^2$ terms
in the large scale expansion of the Green function (\ref{subleadinggreen_app}), noting
that $\gamma_\sigma$ may be written as (\ref{gam}) and comparing to 
(\ref{leadingsource_app}) we see that $k^2 J_k$ contains terms which are of the same size as
those in $J_k$, on large scales.  Thus a consistent large scale expansion of 
$\zeta^\subt_\sigma$ requires that we work
up to order $k^2$ in the expansion of the Green function.

The third contribution to the last line of (\ref{subleadinggreen_app}) is
\begin{eqnarray}
 &-& \frac{\phi'^\subt_k}{\epsilon\sH} - \frac{\phi^\subt_k}{\epsilon} 
  - \frac{k^2\phi^\subt_k}{3\epsilon\sH^2} \ni
  -\frac{5}{6\epsilon}\int_{\tau_i}^{\tau} d\tau'
 \frac{k^2J_k^\sigma}{\sH(\tau')\sH(\tau)^2} \nonumber \\
 &\cong& \frac{5}{6\epsilon}\left[\frac{\gamma'_\sigma}{\sH^3}\right] 
  + \cdots \label{third_app} 
\end{eqnarray}
using the same procedure as above.

Finally, the fourth contribution to the last line of (\ref{subleadinggreen_app}) is
\begin{eqnarray}
 &-& \frac{\phi'^\subt_k}{\epsilon\sH} - \frac{\phi^\subt_k}{\epsilon} 
  - \frac{k^2\phi^\subt_k}{3\epsilon\sH^2} \ni
 \frac{2}{3\epsilon}\int_{\tau_i}^{\tau} d\tau'
 \frac{k^2J_k^\sigma}{\sH(\tau)^3} \nonumber \\
 &\cong& -\frac{2}{3\epsilon}\frac{1}{\sH(\tau)^3}
 \int_{\tau_i}^{\tau}d\tau'\sH(\tau')^2 \Upsilon_3^\sigma \nonumber \\
 &+& \frac{1}{\epsilon}\left[2\frac{\Linv\Upsilon_2^\sigma}{\sH}
 + \frac{2}{3}\frac{\gamma_\sigma}{\sH^2} - \frac{2}{3}\frac{\gamma'_\sigma}{\sH^3}\right]
 + \cdots \label{fourth_app}
\end{eqnarray}
Summing up (\ref{first_app}), (\ref{second_app}), (\ref{third_app}) and (\ref{fourth_app}) and 
inserting the result into (\ref{zeta2_tachyon}) gives
\begin{eqnarray*}
  \zeta_\sigma^\subt &\cong& \frac{1}{\epsilon}\int_{\tau_i}^{\tau}d\tau'
  \left[ - \frac{\Upsilon_1^\sigma}{\sH(\tau')} + 
  \frac{1}{3}\frac{\Upsilon_3^\sigma}{\sH(\tau')}\right.
  \\
  &-& \left. \frac{2}{3}\frac{\sH(\tau')^2}{\sH(\tau)^3}\Upsilon_3^{\sigma}\right]
\end{eqnarray*}
Now, using equations (\ref{Upsilon1sigma}) and (\ref{Upsilon3sigma}) we can write this in 
terms of the tachyon fluctuation $\done\sigma$ as
\begin{eqnarray}
  \zeta^\subt_{\sigma} &\cong& \frac{\kappa^2}{\epsilon}\int_{-1/a_iH}^{\tau}d\tau' \left[
                               \frac{\left(\done\sigma'\right)^2}{\sH(\tau')} 
 - \frac{\sH(\tau')^2}{\sH(\tau)^3}\left( \left(\done\sigma'\right)^2 \right. \right. 
   \nonumber \\
&-& \left. \left. a^2 m_{\sigma}^2\left(\done\sigma\right)^2\right) \right]
\label{final_app}
\end{eqnarray}  

%%%%%%%%%%%%%%%%%%%%%%%%%%%%%%%%%%%%%%%%%%%%%%%%%%%%%%%%%%%%%%%%%%%%%%%%%%%%%%%%%%%%%%%%%%%%

\renewcommand{\theequation}{F-\arabic{equation}}
\setcounter{equation}{0}
\section*{APPENDIX F: The Adiabatic Approximation}
\label{adiabat}

Here we consider the construction of curvature perturbation in the case in 
which the tachyon mass varies adiabatically and demonstrate explicitly that 
the resulting bispectrum is scale invariant.  If the inflaton rolls slowly
enough then the tachyon mass will evolve in time extremely slowly and will
remain close to zero for a significant number of e-foldings.  We define the
slow-roll parameter for the $\sigma$ field as
\beq
\label{etas1}
	\eta_\sigma = 4 M_p^2{m^2_\sigma\over \lambda v^4} = 
	-8 \eta \left({M_p\over v}\right)^2 N
\eeq
For example, if $N_*\sim 30$, then $|\eta_\sigma|$ has the same value at
the beginning and end of inflation.  
We will denote the magnitude of the value of $\eta_\sigma$ at
the end of inflation (and preheating), $t=t_f$, by
\beq
	\eta_f = |\eta_\sigma(t_f)|
\eeq
The condition to have a slowly
rolling tachyon during $2N_*$ e-foldings of inflation is therefore
\beqa
\label{etas}
	\eta_f &<& 1\nonumber\\
 \Longrightarrow\  
	{m^2_\varphi} &<& {\lambda v^6\over 32 N_* M_p^4}
\eeqa
Combining this with the condition (\ref{cobe}), it follows that the coupling
$g$ must be extremely small if $N_*\gsim 1$,
\beq
\label{smallg}
	g < {10^{-4}\over N_*}\,{v\over M_p}
\eeq
Note that a small coupling $g$ does not require fine tuning in the technical
sense, since $g^2$ is only multiplicatively renormalized: $\beta(g^2) \sim
O(g^2\lambda,\ g^4)/(16\pi^2)$.  That is, if $g$ is small at tree
level, loop corrections will not change its effective value 
significantly. The bound (\ref{smallg}) insures that the tachyon will remain light
compared to the Hubble scale during $N_*$ e-foldings of inflation.

The combination of (\ref{smallg}) and (\ref{glb}), eliminating $g$, 
 gives an upper limit
the tachyonic mass scale required for the consistency of our
assumptions, 
\beq
\label{tub}
	\lambda v^2 < {4\times 10^{-7}\over N_*} \, M_p^2
\eeq

We now turn to the adiabatic approximation for the tachyon mode functions.
In this limit the
tachyon mass is varying adiabatically, so we can determine the
behavior of the fluctuations from the deSitter space solutions
by replacing the dependence on a constant tachyon mass with
time-varying one.  The deSitter solutions for massive fields
are given in appendix A.  We are interested in the long-wavelength
growing modes, which  behave like
\beq
\label{appquantum}
  \done \sigma(x) = \int {d^{\,3}k\over (2\pi)^{3/2}}
{H\over\sqrt{2k^3}} (-k\tau)^{\eta_\sigma}
	e^{ikx}\,a_k + {\rm h.c.}
\eeq
in conformal time.  
Since $\eta_\sigma$ is slowly varying, we
can replace it with its time-dependent value (\ref{etas1}), that is
\beq
\label{etas2}	
	{\eta_\sigma}(\tau) = 8\eta\, {M_p^2\over v^2}\,
	\ln {a_*\over a(\tau)}
\eeq
Note that $\ln (a_*/ a(\tau_f)) = -N_*$ at the end of inflation,
so $\eta_f = 8\eta N_* M_p^2/v^2$.
We can use (\ref{end}), which is the criterion for the end of
tachyonic preheating and inflation, 
 to determine $N_*$ in terms of the parameters of the 
hybrid inflation model.
Using (\ref{appquantum}), the
dispersion can be computed at the end of inflation, $\tau=\tau_f$
\beqa
\label{dispersion}
\langle\done \sigma^2(x)\rangle &=& {H^2\over 4\pi^2}
	 \int_{k_0}^{k_{f}} 
{dk\over k} (-k\tau)^{2\eta_\sigma}\\	
	&=& {H^2\over 8\pi^2 \eta_f} \left(
|k_0\tau_f|^{-2\eta_f} - |k_f\tau_f|^{-2\eta_f}\right),
\nonumber
\eeqa
where $\eta_f = |\eta_\sigma(\tau_f)| > 0$.  The infrared and
ultraviolet 
cutoffs
are defined in the same way as is needed to make the dispersion of
the scale-invariant inflaton fluctuations finite:\footnote{The
first order inflaton contribution to the curvature perturbation's
two-point function has the form $\langle(\zeta^{(1)}_\varphi(x))^2\rangle
\sim \int d^{\,3}k/k^3$, and also requires cutoffs}
 $|k_0\tau_i| = 1$ at horizon crossing of
the largest scale modes visible today, so that $|k_0\tau_f| = e^{-N_e}$
with $N_e\sim 60$ being the total number of e-foldings of the visible
part of inflation.  Similarly $|k_f\tau_f|=1$ at the end of
inflation: $k_f$ represents the last mode to cross the horizon
before the end of inflation.  
  Even though the integral converges in the absence of the
cutoff, later on we will need to know the correct behavior of
(\ref{dispersion}) when $\eta_f$ is small, which is not correctly
represented by the limit $k_{\rm max}\to\infty$.  

Eq.\ (\ref{dispersion}) can only describe the growth of the
tachyonic fluctuations until eq.\ (\ref{end}) is fulfilled, and the
perturbative description breaks down.  This signals the end of 
inflation, and allows us to determine the number of e-foldings during
preheating, $N_*$, in terms of the fundamental parameters of the
theory.  Using (\ref{dispersion}) and (\ref{etas1}), the implicit
relation
\beq
	N_* \cong {\lambda v^6\over 64 M_p^4 m_\phi^2}\ln
 \left[1+ 768\pi^2 {N_*} \, {M_p^6 m_\varphi^2\over \lambda^2 v^8}\right]
\eeq
follows.  Using the COBE normalization (\ref{cobe}) to eliminate
$m_\varphi$, this can be rewritten in the form
\beq
\label{nstar}
	N_* \cong {\sqrt{\lambda} v/ M_p\over 15000\, N_e\,
g\sqrt{\lambda}} \ln\left[ 1 + 2\times 10^6\,{N_*g\sqrt{\lambda}}
\left({M_p\over\sqrt{\lambda}v}\right)^{3}\right]
\eeq
This can be solved iteratively on the computer; using an arbitrary
guess for $N_*$ and repeatedly evaluating (\ref{nstar}) gives a
rapidly converging answer.

For consistency, we should check whether we have the freedom to
make $N_*\gsim 1$ given other restrictions on the parameters.
From (\ref{glb}), $g\sqrt{\lambda} > 230 (\sqrt{\lambda}v/M_p)^{3}$,
and taking $N_e\sim 60$, we get
an upper limit on $N_*$,
\beq
	N_* \lsim 10^{-7}\, {M_p^2\over \lambda v^2}\,
	\left(1 + 0.05\,\ln{N_*}\right)
\eeq
This would pose a problem for our scenario if it was not possible
to take $\lambda v^2$ to be
sufficiently small, say $\lambda v^2 / M_p^2 \lsim 10^{-8}$.  
However such a value is consistent with the restriction 
(\ref{tub}), so there is no new constraint coming from
(\ref{dispersion}).  
This demonstrates that
it is consistent to assume that the tachyon remains light until
the end of inflation, provided that $g$ is sufficiently small, 
eq.\  (\ref{smallg}).

We now perform the time integral in the result (\ref{final})
using the knowledge of how $\done\sigma$ behaves
in the long-wavelength limit which will be of interest for
cosmological observables, (\ref{appquantum}).  According to this
result, $\done\sigma'\sim (\eta_\sigma/\tau)\done\sigma$, whereas
$m^2_\sigma a^2 \cong m^2_\sigma /(H^2\tau^2) = 3\eta_\sigma$. 
Therefore, since we are working in the regime where the tachyon is
light, $\eta_\sigma < 1$, 
the term with $m^2_\sigma$ in (\ref{final}) dominates over
the terms with time derivatives.   Let $\tau_f$ be the final time
in the upper limit of integration in (\ref{final}), and $\tau_*$
the time at which the tachyonic instability begins.   For
concreteness, we will at first assume
that there are equal numbers of e-foldings ($N_*\sim 30$) before and
after this time, so that the slow-roll parameter for the tachyon 
is time-antisymmetric, $\eta_\sigma(\tau_i) = - \eta_\sigma(\tau_f)$,
since this is the way of having a light tachyon during
inflation which minimizes the tuning of the tachyon mass.  
Changing variables to $x= - \ln(\tau'/\tau_*)$, and
writing $\eta_\sigma = -d x$ with $d = \eta M_p^2/v^2$, 
eq.\ (\ref{final}) in Fourier space takes the form
\beqa
\label{int}
&& \zeta^\subt_{\sigma}(k) \cong 3 d \frac{\kappa^2}{\epsilon} 
\int{d^{\,3}p\over (2\pi)^{3/2}} \,(\done\hat\sigma)_p\,
(\done\hat\sigma)_{k-p} \\
&&
 \int_{-N_*}^{N_*} dx\, |p\tau_*|^{-dx} |(k-p)\tau_*|^{-dx}
e^{2dx^2 + 3(x-N_*)} + O(d^2) \nonumber
\eeqa
where $\done\hat\sigma_p$ is the part of $\done\tilde\sigma_p$ left after
dividing by the only time-dependent factor, $(-p\tau)^{\eta_\sigma}$.
In fact $\done\hat\sigma\sim H/(2p^3)^{1/2}$ is the perfectly scale-invariant part
of the perturbation, as far as its contribution to the power spectra
of fluctuations,  while the  $(-p\tau)^{\eta_\sigma}$ gives rise to
the small departures from scale invariance.

The integral (\ref{int}) is dominated by contributions near the
final time, so one expects it to behave roughly like the value of
the integrand at $x=N_*$.  By numerically evaluating (\ref{int})
and studying its dependences on $N_*,\ d$ and $k\tau_*$, we are able
to deduce the following semi-analytic fit:
\beqa
\label{int2}
	 \zeta^\subt_{\sigma} &\cong& \frac{\kappa^2}{\epsilon}\,
{\eta_f\over 1 + \eta_f}\, e^{2\eta_f N_*}\, 
\int{d^{\,3}p\over (2\pi)^{3/2}} \,(\done\hat\sigma)_p\,
(\done\hat\sigma)_{k-p} \nonumber\\
&&|p\tau_*|^{-\eta_f}  \, |(k-p)\tau_*|^{-\eta_f} 
\eeqa
where $\eta_f = |\eta_\sigma(\tau_f)|$ is the magnitude of the final
value of $\eta_\sigma$.  Notice that $|p\tau_*| = p/(a_* H)$, so that
a mode which crossed the horizon at the beginning of inflation
(satisfying $|p\tau_i| = 1$) would have $|p\tau_*| = e^{-N_*}$.
The second order perturbation at these scales therefore grows by
a factor of $e^{4\eta_f N_*}$ relative to scale-invariant
perturbations, assuming $|k-p|\sim p$.  This is the growth which is due to the tachyonic
instability.  We remind the reader that this growth saturates 
and our perturbative treatment breaks down when
$\done\sigma \sim v/2$, whereas initially $\done\sigma \sim H$, so
the result (\ref{int2}) is only valid when $e^{4\eta_f N_*}\lsim
v/H$.
	
We can also consider a different case, when horizon crossing occurs
at the time $\tau_*$, and then $N_*\sim 60$ instead of 30.  We find
that the formula (\ref{int2}) is essentially unmodified, because of
the fact that the integral is dominated by its late time behavior,
so that the contributions from the period $\tau<\tau_*$ are
negligible.  In this case, the total growth in $\zeta^\subt_{\sigma}$
comes from the factor $e^{2\eta_f N_*}$ and not from $|k\tau_*|$,
since the latter is of order unity.  Nevertheless the total amount of
growth in the two cases is the same, $\sim (e^{60\eta_f })^2$, and
its range of applicability is limited by the same factor $(v/H)^2$.

We now calculate the leading contribution to the bispectrum of the 
second-order curvature perturbation (\ref{bispectrum}).
Since there is no
momentum dependence in the result (\ref{int2}) for $\zeta^{(2)}$,
this three-point function is straightforward to compute, using the
massless free-field two-point functions
\beq
\left\langle \done\hat{\sigma}_{p_i} \done\hat{\sigma}_{q_i} 
\right\rangle = 
{H^2\over {\sqrt{2|p_i|^3\,2|q_i|^3}}}\,\delta^{(3)}(p_i + q_i) 
\eeq

Carrying out the contractions of pairs of fields which contribute
to the connected part of the bispectrum, one finds eight terms,
which are compactly expressed in terms of the wave numbers $p_i$
and $q_i = k_i - p_i$:
\beqa
&&\left\langle \zeta^{(2)}_{k_1}\, \zeta^{(2)}_{k_2}\,
\zeta^{(2)}_{k_3}\, \right\rangle_{\rm con} = 
(2\pi)^{-9/2}\int \prod_i d^3p_i\, f(p_i,k_i)\nonumber\\
&&  \times\, \Big[ \delta_{p_1+p_2}(\delta_{q_2+p_3}\delta_{q_3+q_1} + 
			 \delta_{q_2+q_3}\delta_{p_3+q_1}) 
\label{bispect}\\ &&+\ 
\delta_{p_1+q_2}(\delta_{p_2+q_3}\delta_{p_3+q_1} + 
	         \delta_{p_2+p_3}\delta_{q_3+q_1})
+(p_i \leftrightarrow q_i) \Big]
\nonumber
\eeqa
where 
\beq
	f(p_i,k_i) = d_* |\tau_*|^{-2\eta_f}
|p_i|^{-3/2-\eta_f}\, |k_i-p_i|^{-3/2-\eta_f}
\eeq 
and $d_* = H^2 \kappa^2 \eta_f  e^{2\eta_f N_*}/(2\epsilon)$.
Each triplet of delta functions contains the factor 
$\delta_{k_1+k_2+k_3}$ expressing the translational invariance of the
bispectrum. The remaining two delta functions can be used to perform
the integrals over $p_1$ and $p_2$, leaving the $p_3\equiv p$ integral.

The resulting bispectrum can be shown (with help from Maple) to reduce to the sum of two terms, 
\beqa
	B(\vec k_i) &=& 4|\tau_*|^{-6\eta_f} d_*^3\int
	{d^{\,3}p\over (2\pi)^3}\,{|p|^{-3-2\eta_f}\,
	|p-k_3|^{-3-2\eta_f}}\nonumber\\
	&&\left({|p+k_2|^{-3-2\eta_f}} + 
	{|p+k_1|^{-3-2\eta_f}}\right)
\eeqa
This expression is manifestly symmetric under $k_1\leftrightarrow
k_2$, but it can also be shown to be symmetric under interchange
of any two $k_i$'s by an appropriate shift of the integration 
variable.  The integral converges as $p\to\infty$, but
since $\eta_f>0$ by definition, it diverges with the small power
$1/p^{\eta_f}$ near $p=0$, and we must introduce an infrared cutoff,
$k_0$, representing a scale a few e-foldings beyond our present
horizon.  This procedure was also necessary for the dispersion of the fluctuations
$\langle(\done\sigma)^2\rangle$ in eq.\ (\ref{dispersion}).  But
unlike in (\ref{dispersion}), we must cut off the integral not just
for $|p|<k_0$, but also for $|p-k_3|<k_0$, $|p+k_1|<k_0$, and
$|p+k_2|<k_0$.  The physical reasoning is however the same: there
should be no observable effects coming from fluctuations whose
wavelength is far beyond our present horizon.

\begin{figure}[htbp]
\bigskip \centerline{\epsfxsize=0.5\textwidth\epsfbox{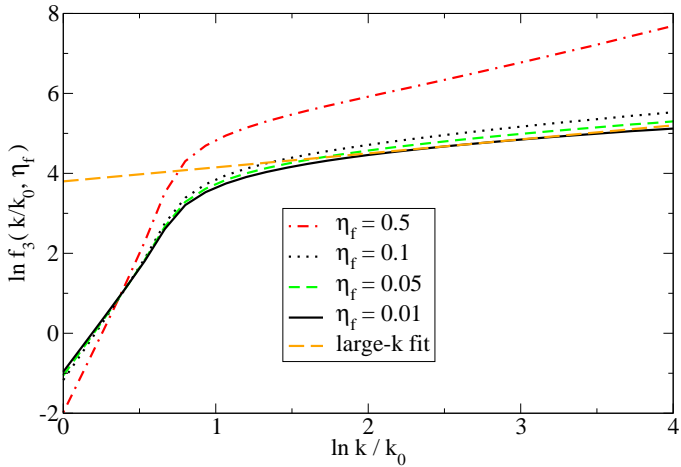}}
%\begin{verse}
%\vskip-0.25cm
\caption{Log of $f_3(k/k_0,\eta_f)$ versus $\ln(k/k_0)$ for several
values of $\eta_f$, and the large-$k$ fit (\ref{fit}).
}
\label{appfig}
\end{figure}

By dimensional analysis, one can see that if all wavenumbers have
the same magnitude, $k_i =  k \hat k_i$, then the bispectrum 
goes like $k^{-6(1+\eta_f)|\tau_*|^{-6\eta_f}}$ times a  
dimensionless function of the directions, $\hat k_i\cdot \hat k_j$,
of $k/k_0$ and of $\eta_f$.
For simplicity, we evaluate it at the symmetric point $\hat k_1 = 
(-1/2,\sqrt{3}/2,0)$, $\hat k_2 = (-1/2,-\sqrt{3}/2,0)$, $\hat k_3 =
(0,1,0)$, where the three wave vectors form an equilateral triangle.
The integral is evaluated numerically to obtain the result
\beqa
\label{appbispectrum2}
	B(\vec k_i) &\cong& {1\over k^6}\, f_3(k/k_0, \eta_f) \, \left({
	H^2 \kappa^2 \eta_f e^{2\eta_f N_*}\over
	2\pi\,\epsilon |k\tau_*|^{2\eta_f}}\right)^3
%,\nonumber\\
%	f(\eta_f) &=& 0.044\, \eta_f^{-1.6 -0.17\ln\eta_f}
\eeqa
We plot the function $f_3(k/k_0, \eta_f)$ in figure \ref{appfig},
which shows strong sensitivity to the presence of the
cutoff for modes with $k\lsim
k_0$, but weak scale-dependence for modes just a few e-foldings 
shorter in wavelength.  Since observable quantities
should not be sensitive to the exact value of the cutoff, 
the region with
$\ln(k/k_0)>1$ is the physically relevant one.  In this region,
a reasonable fit to the result for $\eta_f \lsim 0.1$ is given by
\beq
\label{fit}
	f_3 \sim 45 \left({k\over k_0}\right)^{0.35}
\eeq
which is also plotted in figure \ref{appfig}.  It is clear, then,
that in the case of an adiabatically varying tachyon mass there
is only a small breaking of scale invariance.  This breaking of 
scale invariance can be neglected when evaluating the bispectrum 
near $k=k_0$, corresponding to the largest presently observable scales.  Our
infrared cutoff is effectively at wavelengths a few e-foldings larger
than this scale.  

We have computed the nonlinearity parameter $f_{NL}$ using the approach detailed
in this appendix and have found that large nongaussianity is produced for all values
of the parameters which are consistent with the assumptions we have made.

%%%%%%%%%%%%
%%%%%%%%%%%%

\end{document}